\documentclass[12pt,a4paper]{article}
\pdfoutput=1
\usepackage[british]{babel}
\usepackage[latin1]{inputenc}
\usepackage[usenames,dvipsnames]{color}

\usepackage[intlimits]{amsmath}
\usepackage{amssymb}
\usepackage{aas_macros}
\usepackage{multirow}

\usepackage{url}
\usepackage{graphicx}

\usepackage{cite}
\usepackage{booktabs}

\usepackage[bf,small]{caption}
\usepackage{hyperref}

\hypersetup{
 bookmarksnumbered=true,
 pdftitle = {Antideuterons from Decaying Gravitino Dark Matter}, 
 pdfauthor = {Timur Delahaye, Michael Grefe}, 
 pdfsubject = {antideuterons from decaying gravitino dark matter}, 
 pdfkeywords = {gravitino, dark matter, R-parity violation, cosmic rays, antideuteron, deuteron, antiproton, cross section, coalescence},
 colorlinks=true,
 linkcolor=blue,
 citecolor=blue,
 urlcolor=blue,
}

\numberwithin{equation}{section}

\providecommand{\abs}[1]{\ensuremath{\left\lvert #1\right\rvert }}
\DeclareMathOperator{\BR}{BR}

\begin{document}

\date{\mbox{ }}

\title{
{\normalsize
July 2015\hfill\mbox{}\\}
\vspace{1.5cm}
\bf Antideuterons from \\ Decaying Gravitino Dark Matter\\[8mm]}

\author{Timur Delahaye$^{a,}$\thanks{Electronic address: timur.delahaye@fysik.su.se}\ ~and Michael Grefe$^{b,}$\thanks{Electronic address: michael.grefe@desy.de}\\[2mm]
{\normalsize\it \hspace*{-1cm}$^a$The Oskar Klein Centre for Cosmoparticle Physics, Department of Physics,\hspace*{-1cm}} \\[-2mm] 
{\normalsize\it Stockholm University, AlbaNova University Center, SE-106 91 Stockholm, Sweden} \\
{\normalsize\it $^b$II.\ Institut f\"ur Theoretische Physik, Universit\"at Hamburg,} \\[-2mm]
{\normalsize\it Luruper Chaussee 149, 22761 Hamburg, Germany} 
}
\maketitle

\thispagestyle{empty}

\begin{abstract}
We study the possibility of improving the constraints on the lifetime of gravitino dark matter in scenarios with bilinear $R$-parity violation by estimating the amount of cosmic-ray antideuterons that can be produced in gravitino decays. Taking into account all different sources of theoretical uncertainties, we find that the margin of improvement beyond the limits already set by cosmic-ray antiproton data are quite narrow and unachievable for the next generation of experiments. However, we also identify more promising energy ranges for future experiments.
\end{abstract}

\newpage

\section{Introduction}

Most current cosmic-ray detectors have the ability to determine the mass of the incoming particles -- and in some cases also their charge. However, some isotopes in the cosmic-ray particle spectrum still remain to be observed -- among those antideuterons. Donato, Fornengo and Salati were the first who suggested that antideuterons would be an interesting species for the search of dark matter (DM) via cosmic rays~\cite{Donato:1999gy}. In the past 16 years since this pioneering work, various authors have followed up on this idea~\cite{Baer:2005tw,Donato:2008yx,Brauninger:2009pe,Ibarra:2009tn,Kadastik:2009ts,Cui:2010ud,Grefe:2011kh,Dal:2012my,Ibarra:2012cc,Fornengo:2013osa,Dal:2014nda,Monteux:2014tia}. Indeed, antimatter atoms like antideuterium cannot exist in stars. Therefore, antideuterons -- the corresponding nuclei -- should not exist as astrophysical primary cosmic rays.\footnote{Primary cosmic rays are particles that are accelerated in the ISM, for instance by a supernova remnant.} The only astrophysical process that is expected to produce it is the spallation of cosmic rays off the interstellar medium (ISM). These so-called secondary antideuterons, for kinematic reasons, cannot be produced at kinetic energies lower than few GeV per nucleon~\cite{Chardonnet:1997dv}. If, however, DM annihilations or decays lead to the production of antideuterons, their spectrum would not exhibit this kinematic threshold. This leads to the conclusion that the observed antideuteron flux from DM annihilation or decay in the Galactic halo could be orders of magnitude higher than the astrophysical background at very low energies~\cite{Donato:1999gy}. 

This theoretical argument motivated the construction of dedicated instruments for low-energy antideuteron searches like the General AntiParticle Spectrometer (GAPS)~\cite{Fuke:2008zz}, which after a successful balloon-flight of a prototype in mid-2012 (pGAPS~\cite{Mognet:2013dxa,vonDoetinchem:2013oxa}) is expected to fly again in the future with its final design. Current multi-purpose cosmic-ray experiments, like AMS-02 and BESS, are also looking for this type of particles at somewhat higher energies.

Even though it has been understood that because of tertiary production~\cite{Duperray:2005si}\footnote{Tertiary production refers to non-annihilating inelastic interactions of cosmic-ray antideuterons with the ISM leading to a migration of antideuterons from the high-energy part to the low-energy part of the spectrum.} and energy losses taking place during the cosmic-ray propagation~\cite{Donato:2008yx} the difference between secondary and primary fluxes may not be as high as originally expected, antideuterons are still considered one of the most interesting species for DM searches. More recently, this interest has been extended to other yet unobserved cosmic-ray species like antihelium nuclei from DM annihilations or decays~\cite{Carlson:2014ssa,Cirelli:2014qia}. The growing activity of the field expresses itself through the organisation of dedicated events such as the recent antideuteron workshop held at the University of California, Los Angeles. A summary paper on the status of the field has recently been completed~\cite{Aramaki:2015}.

In the present work, we investigate the case of gravitino DM within the framework of bilinear $R$-parity violation~\cite{Takayama:2000uz,Buchmuller:2007ui}.\footnote{The gravitino could also be a viable DM candidate in scenarios with trilinear $R$-parity violation~\cite{Moreau:2001sr,Lola:2008bk}. Antideuteron signals in this theoretical framework were studied in~\cite{Dal:2014nda,Monteux:2014tia}.} In this type of scenario the gravitino would be long-lived enough to be the DM in the universe but would eventually decay and produce -- among other species -- antideuterons. A main motivation for this scenario is that it leads to a consistent cosmological scenario, explaining the baryon asymmetry of the universe via thermal leptogenesis and avoiding any cosmological gravitino problems~\cite{Buchmuller:2007ui}. For a more detailed introduction of the model, please consult our previous work, where we studied constraints on the gravitino lifetime from cosmic-ray antiprotons in the same theoretical framework~\cite{Delahaye:2013yqa}.

The structure of this paper is as follows: In the next section, we briefly revise deuteron and antideuteron formation in the coalescence model, derive the coalescence momentum from collider data, and simulate antideuteron spectra for the gravitino decay channels. In section 3, after studying the cross sections at stake, we compute the propagated cosmic-ray antideuteron fluxes at Earth, taking into account various sources of uncertainties. In section 4, we discuss the prospects for the detection of gravitino decays via antideuterons, before we come to our conclusions. In an appendix to this paper, we present some cross section parametrisations relevant for antideuteron propagation.

\section{Deuteron and Antideuteron Formation}

\subsection{The Coalescence Model}
Deuterons and antideuterons may form in particle physics processes through final state interactions in a nucleon shower~\cite{Butler:1963pp}. A usual approach to describe their formation is the phenomenological coalescence model~\cite{Schwarzschild:1963zz,Csernai:1986qf}. In this prescription, independent of the details of the microscopic formation mechanism, a deuteron is formed when a proton and a neutron come sufficiently close in momentum space, i.e.\ when the absolute value of the difference of their four-momenta is below a threshold coalescence momentum:
\begin{equation}
  \abs{p_p-p_n}<p_0\,.
\end{equation}
In order to determine the value of the coalescence momentum relevant for antideuteron production in gravitino decays we will make use of the antideuteron production rate measured by the ALEPH experiment at the LEP collider~\cite{Schael:2006fd}. Although there are several measurements of antideuteron production in collider experiments -- which in general do not lead to a common value for $p_0$~\cite{Aramaki:2015,Ibarra:2012cc} -- we restrict to the ALEPH data since the gravitino decay channel $\psi_{3/2}\rightarrow Z\nu$ produces antideuterons via $Z$ boson fragmentation exactly as in the ALEPH measurement. Also the channels $\psi_{3/2}\rightarrow W\ell$ and $\psi_{3/2}\rightarrow h\nu$ are expected to be more similar to the case of $Z$ boson fragmentation than to the production of antideuterons in $pp$ collisions~\cite{Alper:1973my,Henning:1977mt,Sharma:2012zz}, $e^-p$ collisions~\cite{Chekanov:2007mv}, or $e^+e^-$ collisions on or nearby the Upsilon resonance~\cite{Asner:2006pw,Lees:2014iub}.

A common approximation for deuteron formation is that the distributions of neutrons and protons in momentum space are spherically symmetric and uncorrelated~\cite{Butler:1963pp}. This leads to an energy distribution of deuterons that can be directly calculated from the product of the individual spectra of neutrons and protons~\cite{Kadastik:2009ts,Brauninger:2009pe}:
\begin{equation}
  \frac{dN_d}{dT_d}=\frac{p_0^3}{6}\,\frac{m_d}{m_p\,m_n}\,\frac{1}{\sqrt{T_d^2+2\,m_d\,T_d}}\,\frac{dN_p}{dT_p}\,\frac{dN_n}{dT_n}\,,
  \label{factorised}
\end{equation}
where $T_p=T_n=T_d/2$ are the kinetic energies of protons, neutrons and deuterons. Therefore, this approximation is called factorised coalescence.

As pointed out in~\cite{Kadastik:2009ts}, this approximation is qualitatively wrong and significantly underestimates the deuteron yield in high-energetic processes. This is due to the fact that the distributions of neutrons and protons are actually neither spherically symmetric nor uncorrelated. For instance, in the decay of a DM particle into a $Z$ boson and a neutrino the probability for the formation of a deuteron in the fragmentation of the $Z$ boson should be independent of the DM mass. This is due to the fact that the $Z$ boson fragmentation process is always the same as viewed from the $Z$ boson rest frame. However, the factorised coalescence approximation gives a lower yield of deuterons for larger DM masses, since the protons and neutrons are distributed over a larger phase space for higher injection energies. Another qualitatively wrong behaviour of this approximation is the possibility that protons and neutrons from distinct DM decays form a deuteron. This is due to the fact that the spectra of protons and neutrons are simply multiplied, while in principle the coalescence condition on the four-momenta of protons and neutrons should be applied on an event-by-event basis.

This can be achieved in a Monte Carlo simulation of the decay process~\cite{Kadastik:2009ts}. For instance, using an event generator like \textsc{Pythia} 6.4~\cite{Sjostrand:2006za} one can simulate the hadronisation of massive gauge and Higgs bosons and then apply, event by event, the coalescence condition on the protons and neutrons.\footnote{In fact, there are some combinatoric ambiguities in the deuteron coalescence in \textsc{Pythia}. Namely, it is possible that a proton fulfils the coalescence criteria with two different neutrons or vice versa. We checked numerically that the number of events where this is the case is very low. The probability increases with $p_0$, but even for $p_0=250\,$MeV less than 0.1\,\% of the deuteron events are affected. Therefore, we conclude this not to be a significant effect. In our analysis we always combine the first pair of protons and neutrons fulfilling the coalescence criterion according to the particle order in the event record.} This method leads to plausible results, e.g.\ the deuteron yield in the DM decay to final states including $W$, $Z$ or Higgs bosons is independent of the DM mass. However, this strategy also requires a lot of computing time to generate smooth spectra as only one deuteron or antideuteron is produced in $\mathcal{O}(10^4)$ fragmentation processes. Moreover, the authors of~\cite{Artoisenet:2010uu} point out that baryon pair distributions are not used to tune event generators and therefore their predictions for deuteron production are not without uncertainty. In fact, the authors of~\cite{Dal:2012my} find a discrepancy of up to a factor 2--4 when comparing deuteron production between the Monte Carlo generators Herwig++~\cite{Bahr:2008pv,Gieseke:2011na} and \textsc{Pythia} 8~\cite{Sjostrand:2007gs}.

Due to the deuteron's binding energy of 2.224\,MeV~\cite{Agashe:2014kda}, the coalescence of free protons and neutrons is forbidden since energy and momentum cannot be conserved at the same time if no further particles are involved in the process. 
However, in a hadronic shower with many particles, 
it is easily conceivable that the overall process conserves energy and momentum and thus allows for deuteron coalescence.\footnote{While low-energy deuteron formation proceeds via the process $p\,n\rightarrow d\,\gamma$, this is not necessarily the case in high-energy nuclear collisions. In their pioneering work on deuteron formation, Butler and Pearson considered an explicit interaction with the optical potential of the target nucleus to take care of energy and momentum conservation~\cite{Butler:1963pp}. However, this model was later found to be in conflict with experimental data~\cite{Mrowczynski:1987,Kolybasov:1989}. The factorised coalescence model does not touch the issue at all since it does not go into the details of the microscopic formation mechanism~\cite{Csernai:1986qf}. In the 1980s it was argued that no explicit interaction with a third body is needed since deuterons are formed in a tiny space-time region and thus the uncertainty principle applies~\cite{Mrowczynski:1987}, or that the dominant process for deuteron formation in a hadronic shower involves a proton and a neutron that are slightly off-shell~\cite{Kolybasov:1989,Duperray:2002pj}. Nonetheless, other authors still stressed the necessity of an explicit interaction with a third body in the nuclear shower~\cite{Leupold:1994,Scheibl:1998tk}. The authors of~\cite{Dal:2015sha} recently discussed an alternative deuteron formation model that makes use of measured cross sections for deuteron production processes with photon or pion emission.} When simulating deuteron coalescence in a Monte Carlo event generator, the four-momenta of protons and neutrons are given explicitly and we have no way of realistically treating the multi-body process leading to deuteron formation. The most common approach in previous works on antideuteron signals from DM annihilation or decay is thus to consider momentum conservation and to determine the deuteron kinetic energy from the proton and neutron momenta and the deuteron mass of 1.8756\,GeV~\cite{Agashe:2014kda}:
\begin{equation}
  T_d=\sqrt{m_d^2+(\vec{p}_p+\vec{p}_n)^2}-m_d\,.
\end{equation}
We will also follow this approach since it turns out that the uncertainty on the spectra introduced by this ambiguity is marginal for our discussion.\footnote{We checked how alternative approaches affect the resulting deuteron spectra. Starting from energy conservation, i.e.\ determining the deuteron kinetic energy from the proton and neutron energies and the deuteron mass to be $T_d=E_p+E_n-m_d$, the resulting spectra differ only notably at deuteron momenta below the coalescence momentum.}

More recently, it was realised that the condition $\abs{p_p-p_n}<p_0$ is not not enough to guarantee a physically meaningful coalescence prescription in Monte Carlo event generators~\cite{Ibarra:2012cc}. Some of the antiprotons and antineutrons emerging in DM annihilations or decays are generated by the decay of metastable mother particles like the baryons $\bar{\Lambda}^0$, $\bar{\Sigma}^\mp$, $\bar{\Lambda}_c^-$, $\bar{\Lambda}_b^0$, $\bar{\Xi}_b^{0,\pm}$ and $\Omega_b^-$, and the mesons $D_s^\pm$, $B^{0,\pm}$ and $B_s^0$, which have lifetimes of $\mathcal{O}(10^{-10}$--$10^{-13})\,$s~\cite{Sjostrand:2006za,Agashe:2014kda}. By contrast, the nuclear interactions leading to antideuteron formation take place on femtometer length scales, corresponding to lifetimes of $\mathcal{O}(10^{-23})\,$s for relativistic particles. Therefore, it is necessary to include a condition on the spatial separation of protons and neutrons as well. 

To accommodate this, Ibarra and Wild excluded weakly decaying baryons with lifetimes $\tau > 1$\,mm/c from the decay chain~\cite{Ibarra:2012cc}.\footnote{The exact criterion is not stated in the paper; S.~Wild, private communication (2014).} Although this condition removes the most relevant metastable mother particles, namely $\bar{\Lambda}^0$ and $\bar{\Sigma}^\mp$, several other metastable particles producing antiprotons and antineutrons have lifetimes just below 1\,mm/c. Fornengo \textit{et al.} introduced the condition $\Delta r < 2\,$fm on the spatial separation~\cite{Fornengo:2013osa}. In \textsc{Pythia} 6.4, this treatment is basically equivalent to considering only antiprotons and antineutrons with a production time $t=0$ since only non-zero lifetimes that are relevant for displaced vertices in collider experiments are stored in the event table. This treatment removes all antiprotons and antineutrons coming from metastable mother particles and this is the treatment we will adopt for our analysis below.

\subsection{Determination of the Coalescence Momentum}
\begin{figure}[t]
 \includegraphics[width=0.49\linewidth]{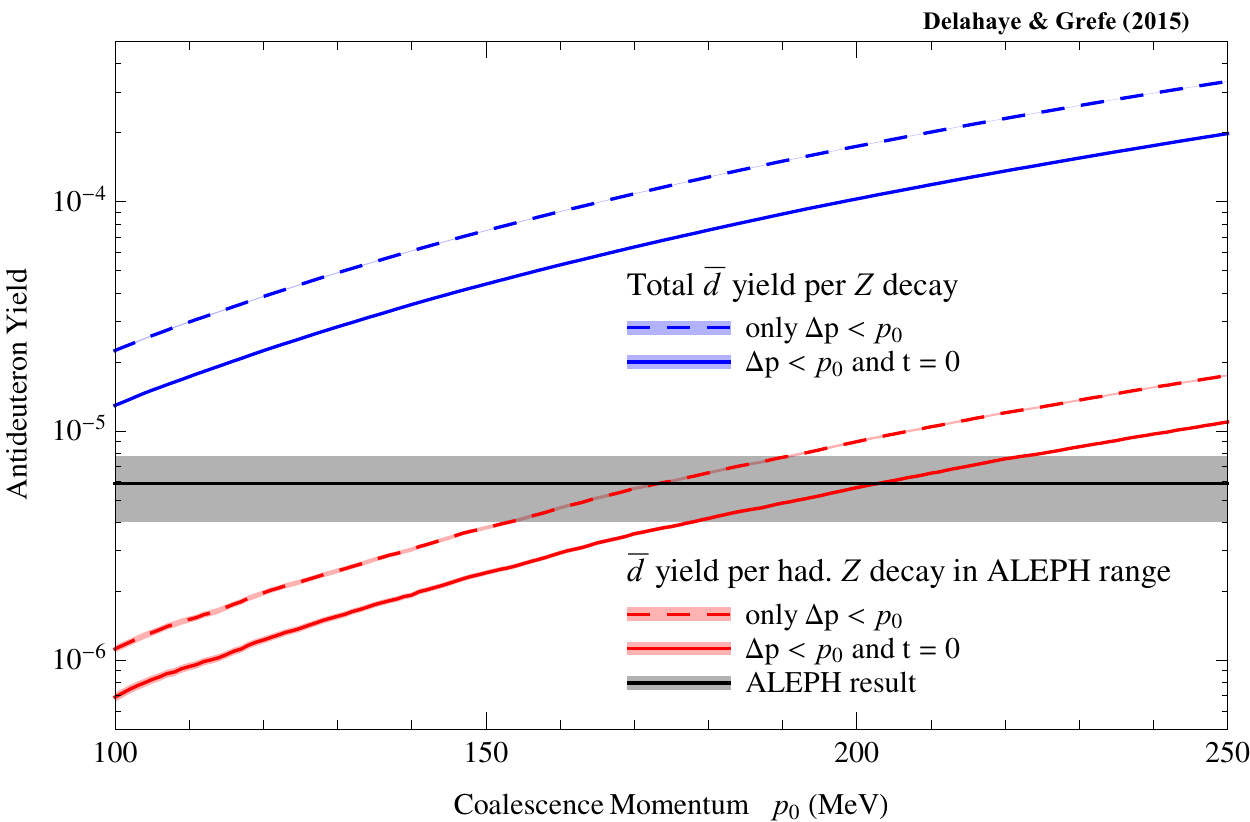} 
 \hfill
 \includegraphics[width=0.50\linewidth]{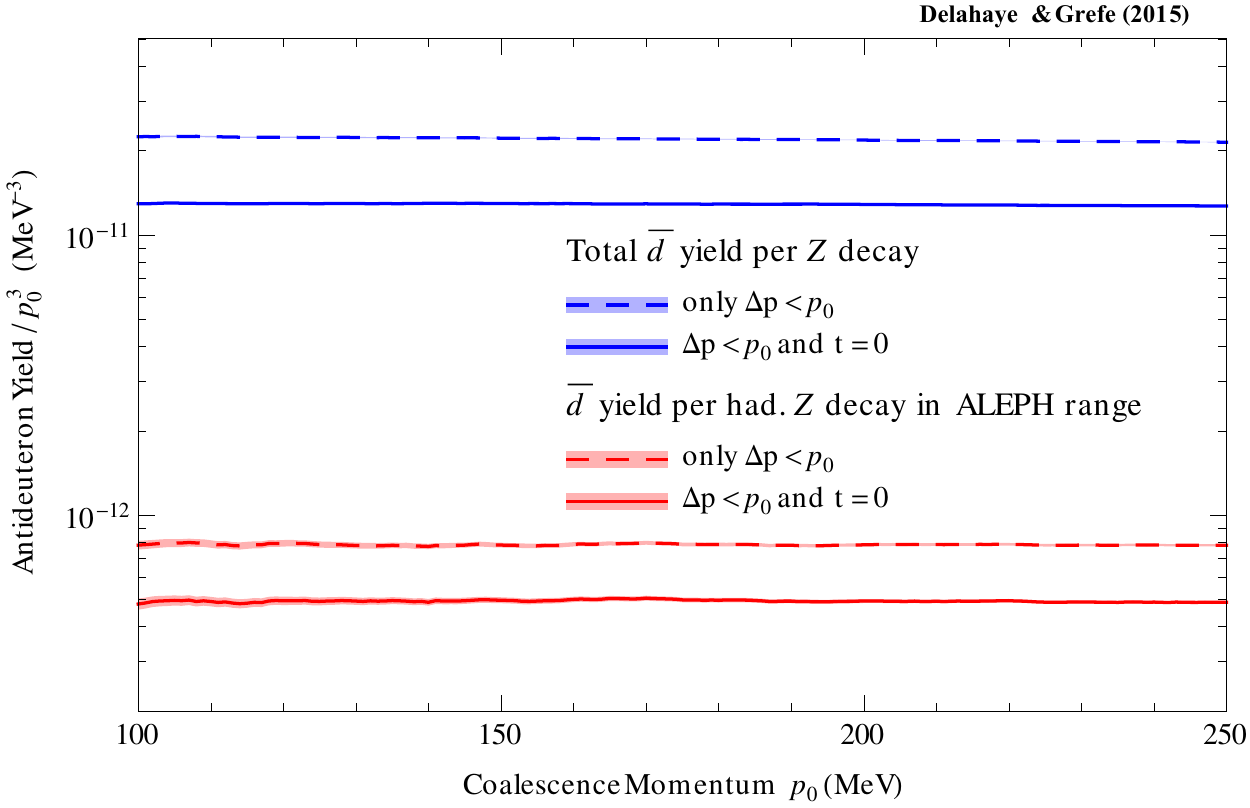} 
 \caption{\textit{Left:} Antideuteron yield as a function of the coalescence momentum $p_0$. The upper (blue) lines show the total antideuteron yield and the lower (red) lines show the antideuteron yield after applying the same cuts as the ALEPH analysis. The ALEPH result is shown for comparison as a grey horizontal band. \textit{Right:} Antideuteron yield divided by $p_0^3$ as a function of $p_0$. It is clearly visible that the antideuteron yields scale with the third power of the coalescence momentum. The solid lines are our main result. The dashed lines show the antideuteron yields without excluding antiprotons and antineutrons coming from metastable mother particles. In all cases the coloured bands show the uncertainty introduced by Monte Carlo statistics.}
 \label{coalescence}
\end{figure}

Using \textsc{Pythia} 6.4, we simulated $10^9$ events of the process $e^+e^- \rightarrow Z$ to determine $p_0$ from ALEPH data. The authors of~\cite{Schael:2006fd} found the number of antideuterons produced per hadronic $Z$ boson decay in the process $e^+e^- \rightarrow Z \rightarrow \bar{d} + X$ at the $Z$ pole to be\footnote{The first error gives the statistical error, while the second error corresponds to systematic errors.}
\begin{equation}
  R_{\bar{d}} = (5.9 \pm 1.8 \pm 0.5) \times 10^{-6}
\end{equation}
within the momentum range $0.62\,\text{GeV} < p_{\bar{d}} < 1.03\,\text{GeV}$ and in the angular range $|\cos{\theta}|<0.95$. In the left panel of figure~\ref{coalescence} we present the simulated antideuteron yield as a function of the coalescence momentum. We show the total antideuteron yield per $Z$ decay event as well as the yield per hadronic $Z$ decay applying the momentum and angular cuts of the ALEPH analysis. The hadronic branching ratio of the $Z$ boson is 69.91\,\%. Leptonic $Z$ decays do not contribute at all to antideuteron production~\cite{Agashe:2014kda}. We overlay the ALEPH result, adding statistical and systematic errors in quadrature. The best-fit values for the coalescence momentum and the 1-$\sigma$ ranges can be basically read off this plot and are summarised in table~\ref{coalescencemomentum} along with the value following from the factorised coalescence model, see eq.~(\ref{factorised}). In the right plot of figure~\ref{coalescence} we demonstrate that in all cases the dependence of the antideuteron yield on the coalescence momentum is in very good agreement with the expected $p_0^3$ behaviour.

\begin{table}
 \centering
 \begin{tabular}{lc}
  \toprule
  Coalescence Model & Coalescence Momentum $p_0$ \\
  \midrule
  Factorised Coalescence & $141_{-16}^{+14}\,$MeV \\
  \textsc{Pythia} 6.4 (only $\Delta p<p_0$) & $173_{-19}^{+18}\,$MeV \\
  \textsc{Pythia} 6.4 ($\Delta p<p_0$ and $t=0$) & $203_{-25}^{+20}\,$MeV \\
  \bottomrule
 \end{tabular}
 \caption{Coalescence momenta for antideuteron production determined from the ALEPH data. We compare the factorised coalescence approach using antiproton and antineutron spectra generated with \textsc{Pythia} to the event-by-event coalescence using either only the momentum condition or additionally excluding antiprotons and antineutrons produced in decays of long-lived mother particles.}
 \label{coalescencemomentum}
\end{table}

\begin{figure}[t]
 \includegraphics[width=0.49\linewidth]{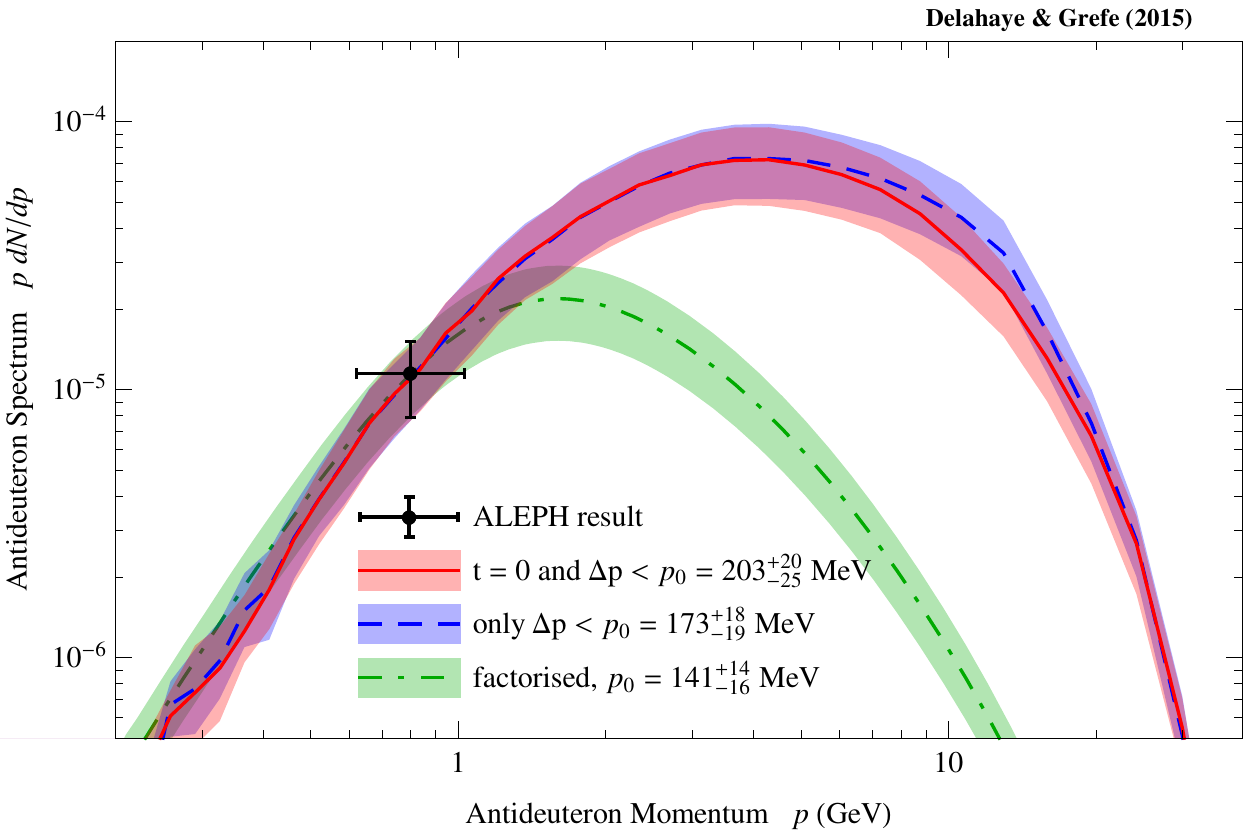} 
 \hfill
 \includegraphics[width=0.49\linewidth]{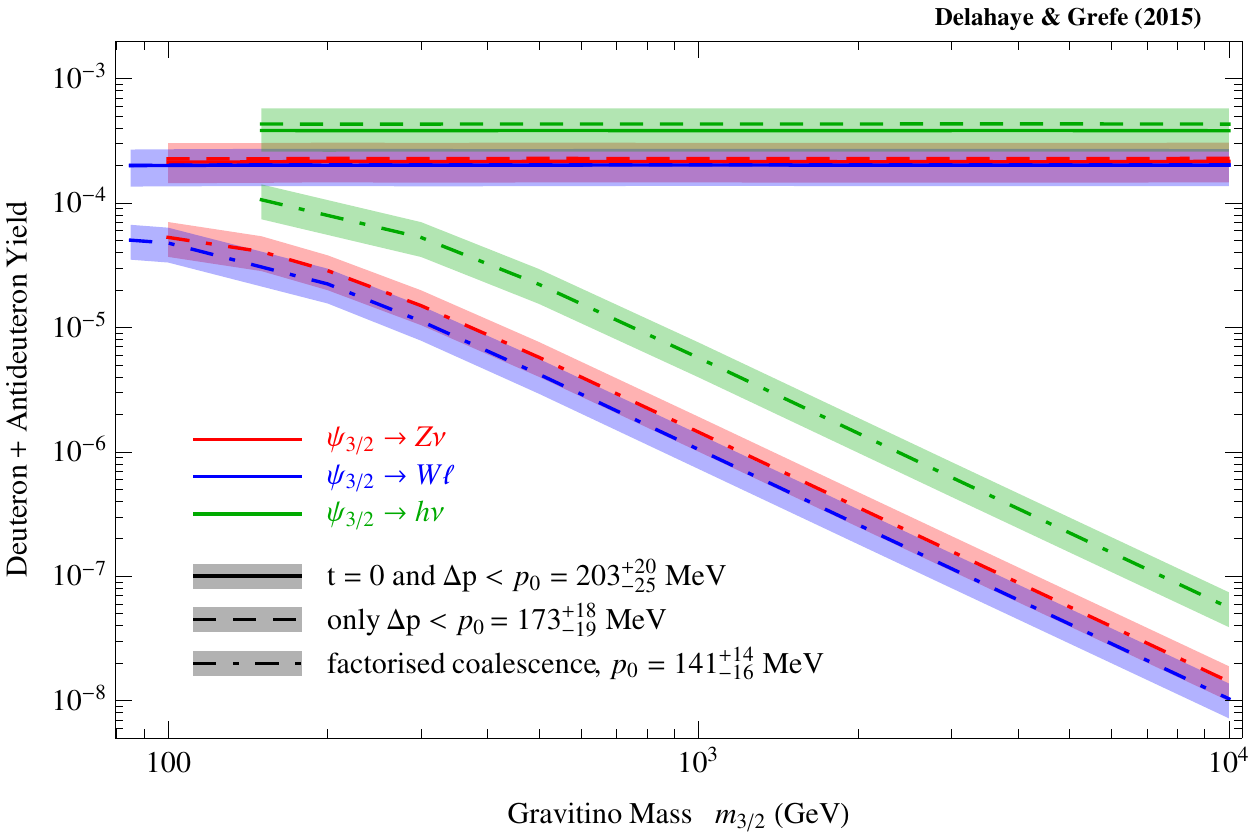} 
 \caption{\textit{Left:} Simulated antideuteron spectra from the process $e^+e^- \rightarrow Z \rightarrow \bar{d} + X$ compared to the measurement of the ALEPH experiment at LEP. The spectrum obtained from factorised coalescence is shown in green. The spectra derived from \textsc{Pythia} are shown in blue (only $\Delta p<p_0$) and red ($\Delta p<p_0$ and $t=0$). The two Monte Carlo spectra are very similar except for the high-energy part. The spectrum derived from uncorrelated antiprotons and antineutrons exhibits a very different shape. \textit{Right:} Antideuteron yields in the decay channels $Z\nu$, $W\ell$ and $h\nu$ as a function of $m_{3/2}$. The yields obtained from the \textsc{Pythia} simulation are independent of the gravitino mass and very similar to each other. By contrast, the factorised coalescence prescription leads to unphysical behaviour, with yields falling off like $m_{3/2}^{-2}$.}
 \label{ALEPHyield}
\end{figure}
In the left panel of figure~\ref{ALEPHyield}, we present the antideuteron spectrum in $Z$ decays as a function of the momentum for the three different treatments of coalescence we discussed above. The spectrum derived from the factorised coalescence approximation is completely different from the spectra based on event-by-event coalescence. By contrast, the exclusion of antiprotons and antineutrons from metastable mother particles only slightly changes the shape of the high-energy part of the spectrum. The single ALEPH data point cannot give any constraints on the shape of the spectrum. Therefore, all coalescence treatments fit the data equally well and our choice in favour of the Monte Carlo simulation with $\Delta p<p_0$ and $t=0$ is only based on the theoretical considerations discussed above. For the rest of the paper we will stick to the event-by-event coalescence using $p_0 = 203_{-25}^{+20}\,$MeV and excluding antiprotons and antineutrons produced at displaced vertices.

Let us conclude this section with a comparison to other $p_0$ determinations in the literature. Our $p_0$ values differ somewhat from the values found by the authors of~\cite{Kadastik:2009ts},\cite{Ibarra:2012cc} and~\cite{Fornengo:2013osa}. Kadastik \textit{et al.} used \textsc{Pythia}~8 and found $p_0=162\pm17\,$MeV for the $\Delta p<p_0$ condition; roughly 10\,MeV below our result. Ibarra \textit{et al.} also used \textsc{Pythia}~8 and found a coalescence momentum of $p_0=192\pm30\,$MeV when removing long-lived mother particles; again 10\,MeV below our result. Fornengo \textit{et al.} used \textsc{Pythia}~6.4 and found $p_0=195\pm22\,$MeV when excluding metastable mothers. For the $\Delta p<p_0$ condition they found $p_0=180\pm18\,$MeV and for the factorised coalescence approximation they found $p_0=160\pm19$\,MeV.\footnote{After going back to their code, they now find $p_0=143\,$MeV for the latter case, compatible with our result; N.~Fornengo, private communication (2014).} Putze finds $p_0=202\,$MeV when using \textsc{Pythia}~6.4, a value similar to ours, and $p_0=195\,$MeV when using \textsc{Pythia}~8, compatible with Ibarra \textit{et al.}~\cite{Putze:2014}. As mentioned before, differences between different event generators are not completely unexpected. However, in some cases there are also different results for the same tool. These differences are a bit worrisome since the Monte Carlo statistical uncertainty is only of the order of few MeV. A potential source of systematic uncertainty is the choice of the Monte Carlo tune, but to our knowledge all of the studies listed above use the default tune of \textsc{Pythia}~6.4/8.

\subsection{Antideuteron Spectra from Gravitino Decay}

In models with bilinear $R$-parity violation, the gravitino has four main decay channels: $\psi_{3/2}\rightarrow\gamma\nu$, $Z\nu$, $W\ell$ and $h\nu$~\cite{Ishiwata:2008cu,Covi:2008jy}. As for the case of antiprotons~\cite{Delahaye:2013yqa}, only the channels containing a massive gauge or Higgs boson in the final state are relevant for antideuterons. For the generation of the antideuteron spectra from gravitino decay we simulated $10^9$ events with \textsc{Pythia} 6.4 for each of the decay channels and for a set of gravitino masses of roughly equal distance on a logarithmic scale: $m_{3/2}=85\,$GeV, 100\,GeV, 150\,GeV, 200\,GeV, 300\,GeV, 500\,GeV, 1\,TeV, 2\,TeV, 3\,TeV, 5\,TeV, and 10\,TeV. Using the same strategy as in~\cite{Delahaye:2013yqa}, we started the \textsc{Pythia} simulation with a resonance decay into two particles, $Z$ boson and neutrino, $W$ boson and charged lepton, and Higgs boson and neutrino, respectively. In this way the $Z$, $W$ and Higgs bosons are treated as decaying isotropically in their rest frames. 

\begin{figure}[t]
  \centering
  \includegraphics[width=0.325\textwidth]{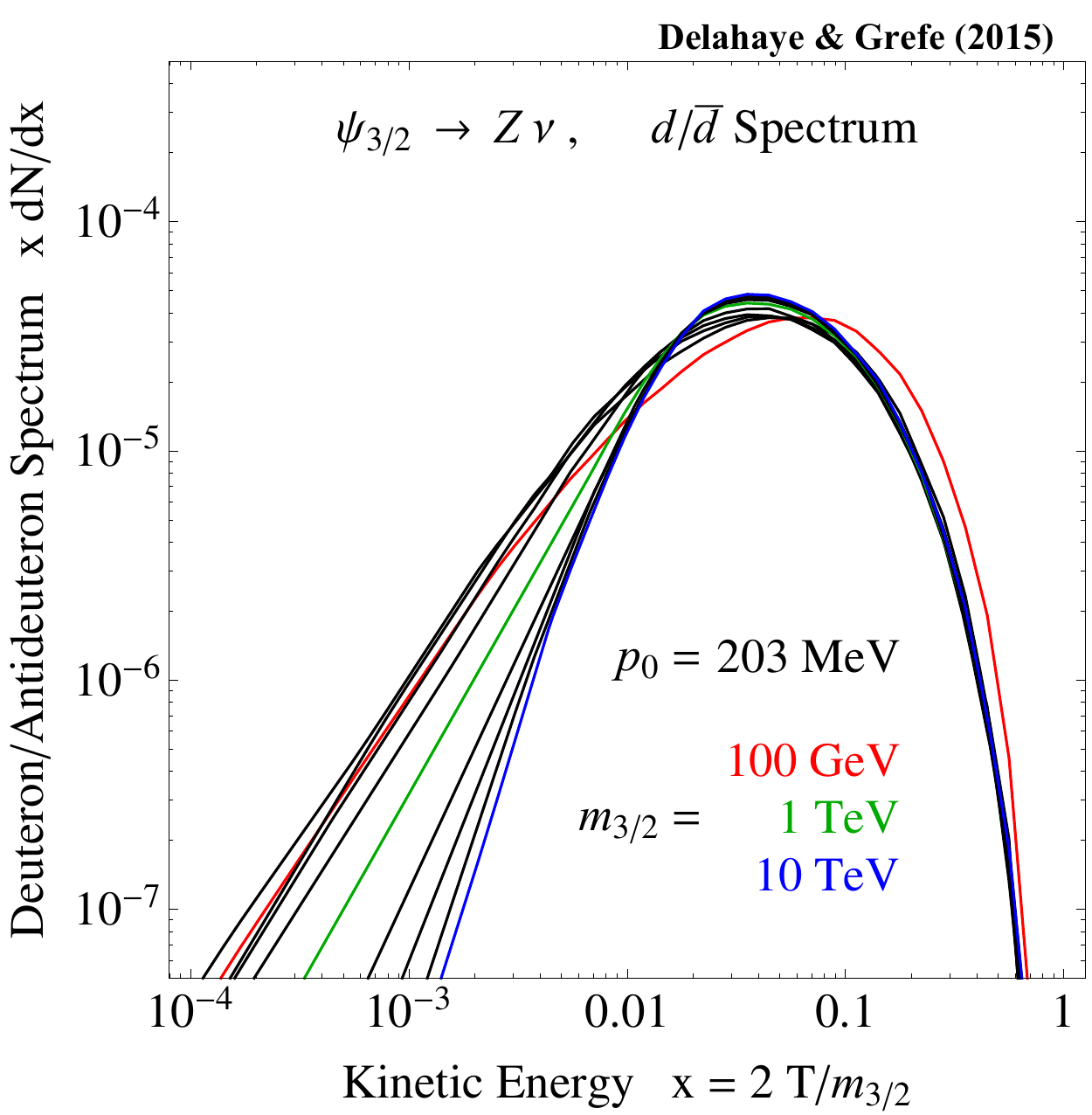} 
  \hfill
  \includegraphics[width=0.325\textwidth]{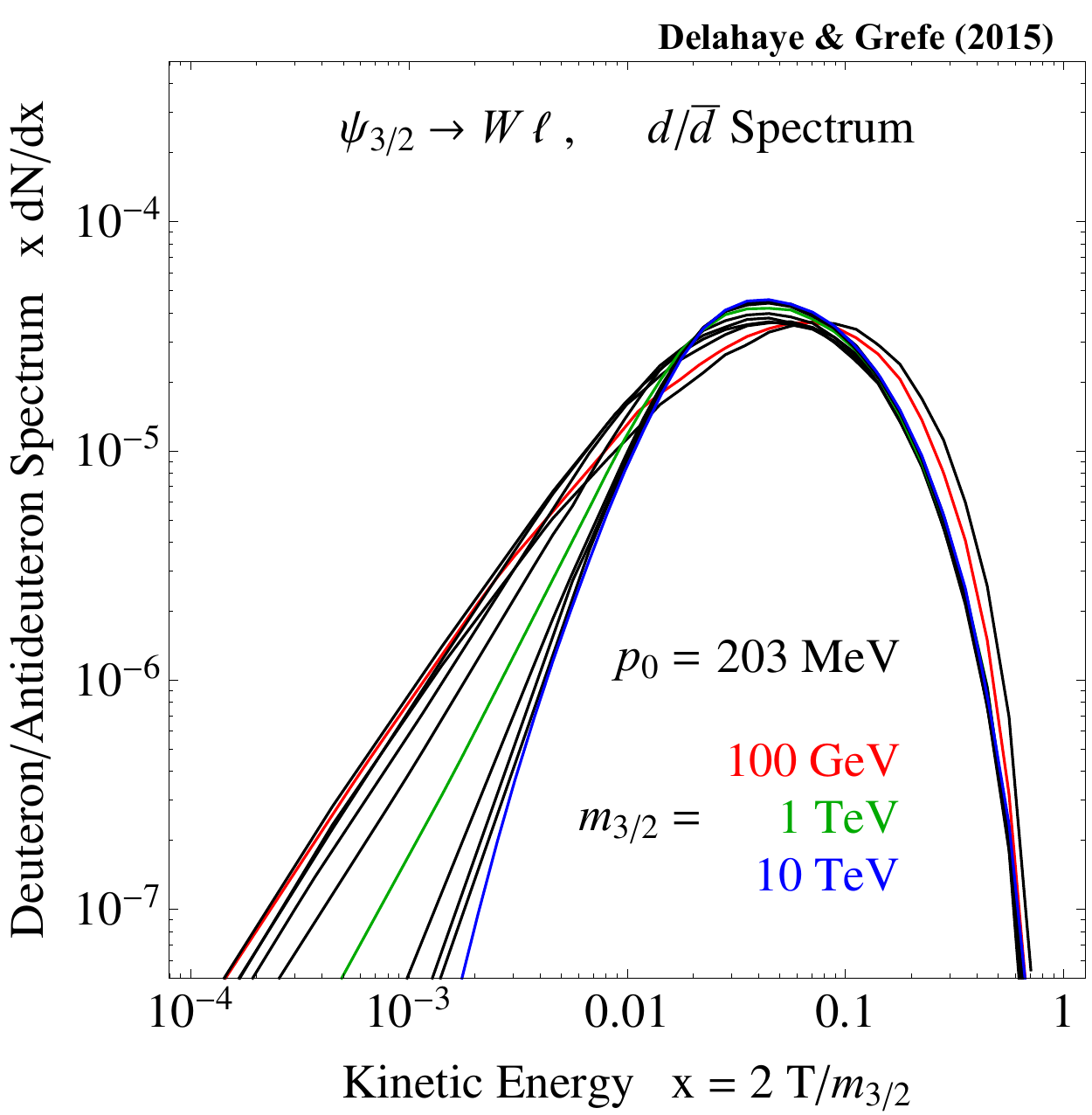} 
  \hfill
  \includegraphics[width=0.325\textwidth]{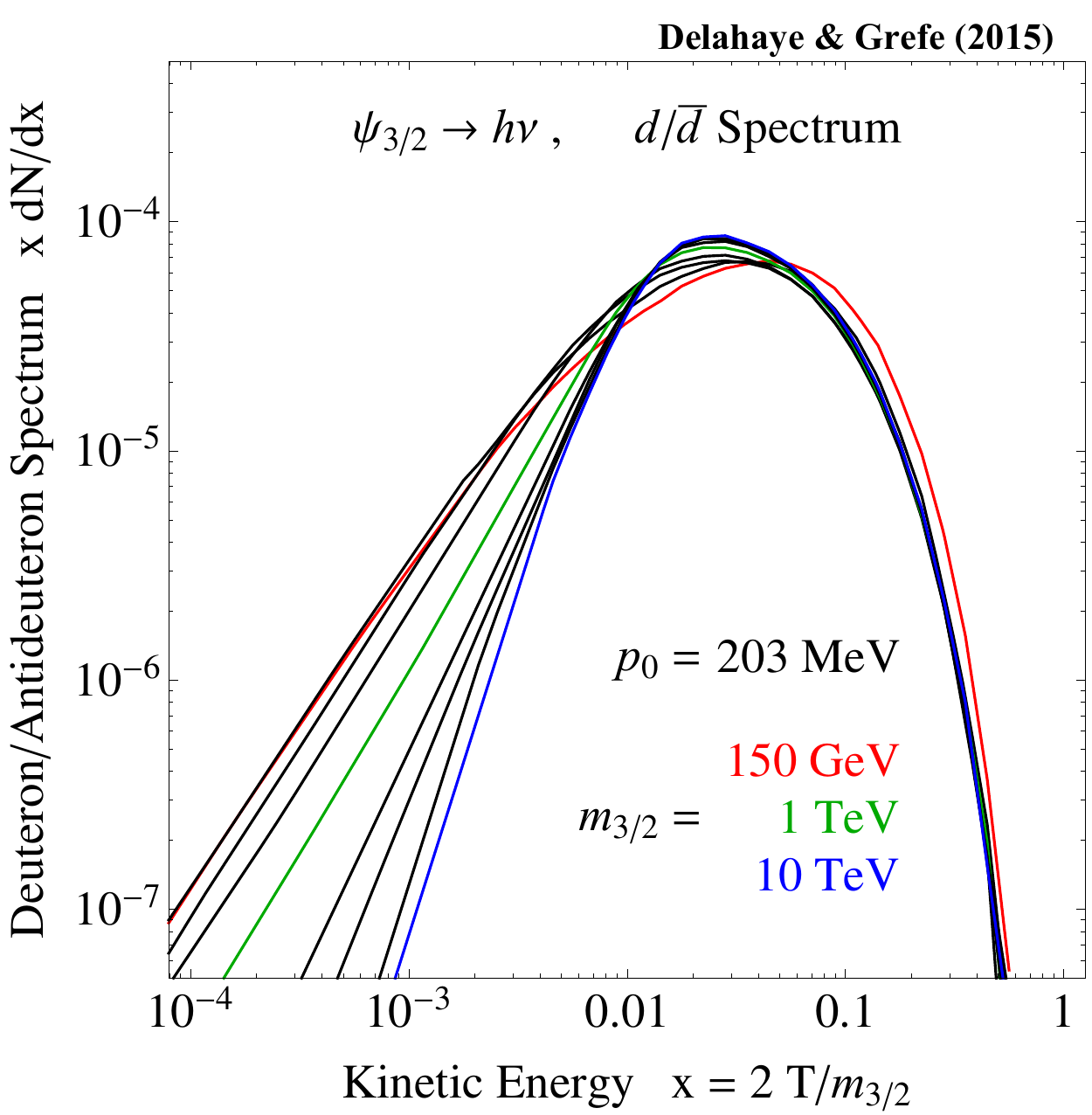} 
  \caption{Deuteron/antideuteron spectra from the two-body decay of a gravitino into $Z\nu$ (\textit{left}), $W\ell$ (\textit{centre}) or $h\nu$ (\textit{right}). The spectra are shown for the central value of $p_0 = 203\,$MeV and for gravitino masses of 150\,GeV (red), 200\,GeV, 300\,GeV, 500\,GeV, 1\,TeV (green), 2\,TeV, 3\,TeV, 5\,TeV, and 10\,TeV (blue). All spectra are normalized to the respective gravitino mass.}
  \label{decayspectra}
\end{figure}

The antideuteron yields in these decay channels are expected to be independent of the gravitino mass since the antideuterons are produced in the fragmentation of an on-shell gauge or Higgs boson. Larger gravitino masses should thus only lead to boosted spectra, while the underlying physical process remains unchanged.\footnote{Note that we did not take into account weak corrections that would lead to a slight increase of the yields with increasing gravitino mass, see for instance~\cite{Ciafaloni:2010ti}.} Our simulation confirms this expectation and shows that the factorised coalescence prescription leads to unphysical results, see the right panel of figure~\ref{ALEPHyield}. The antideuteron yields per gravitino decay are summarised in table~\ref{decaymultiplicities}.\footnote{Note that these results differ from those reported in~\cite{Grefe:2011dp} due to an erroneous treatment of the \textsc{Pythia} routine in the earlier study.} The resulting spectra are presented in figure~\ref{decayspectra} for the central value of $p_0 = 203\,$MeV.\footnote{The decay spectra are available in tabulated form at \url{http://www.desy.de/~mgrefe/files.html}.} Although these spectra by eye appear to have the same shape as the antiproton spectra from gravitino decay (see~\cite{Delahaye:2013yqa}), rescaled by a factor of $\mathcal{O}(10^{-4})$, there is no simple scaling relation among them.
\begin{table}
 \centering
 \begin{tabular}{cccccc}
  \toprule
  Particle type & $Z\nu$ & $W\ell$ & $h\nu$ \\
  \midrule
  $d+\bar{d}$ (t=0) & $(2.15\pm0.69)\times10^{-4}$ & $(2.01\pm0.65)\times10^{-4}$ & $(3.82\pm1.22)\times10^{-4}$ \\
  $d+\bar{d}$ (only p) & $(2.27_{\,-\,0.66}^{\,+\,0.77})\times10^{-4}$ & $(2.02_{\,-\,0.59}^{\,+\,0.69})\times10^{-4}$ & $(4.31_{\,-\,1.24}^{\,+\,1.44})\times10^{-4}$ \\
  \bottomrule
 \end{tabular}
 \caption{Multiplicities of deuterons + antideuterons from gravitino decays after the event-by-event coalescence process simulated with \textsc{Pythia} 6.4. The central values are given for $p_0=203$\,MeV ($t=0$) and $p_0=173$\,MeV ($p<p_0$), respectively. The quoted errors of roughly 30\% correspond to the 1-$\sigma$ uncertainty in the determination of $p_0$.}
 \label{decaymultiplicities}
\end{table}

We will now move on to the discussion of the antideuteron flux observed at Earth. The flux expected from gravitino decay is simply a linear combination of the fluxes for the individual decay channels:
\begin{equation}
 \Phi_{\bar{d}}=\BR(Z\nu)\,\Phi_{\bar{d}}^{Z\nu}+\BR(W\ell)\,\Phi_{\bar{d}}^{W\ell}+\BR(h\nu)\,\Phi_{\bar{d}}^{h\nu}.
\end{equation}
The branching ratios for the different decay channels depend on the choice of the supersymmetry parameters and are discussed in detail in~\cite{Grefe:2011dp,Delahaye:2013yqa}. Note that they do not depend on the amount of $R$-parity violation but only on the mass of the gravitino and the mass hierarchy of the neutralino sector of the supersymmetric particle spectrum. It is hence convenient to label the different cases by the name of the next-to-lightest supersymmetric particle (NLSP). In this work we use the values presented in figure~3 and table~3 of~\cite{Delahaye:2013yqa}. For illustration, we restrict to an example case where the NLSP is Bino-like.

\section{Antideuteron Flux from Gravitino Decay}

As explained in the introduction, antideuterons allow for an almost background-free search for an exotic DM component in certain parameter ranges. In fact, no cosmic-ray antideuterons have been observed so far and there exists only an upper limit on the antideuteron flux from the BESS experiment~\cite{Fuke:2005it}. In addition, BESS-Polar II looked for antideuterons using more than ten times more cosmic-ray data than the previous BESS analysis. No candidate antideuterons were observed and a flux limit is expected to be published in the near future~\cite{Yoshimura:2013,Sasaki:2014}. In addition, several experiments are currently taking data or will start operating within the next years to improve this situation: The AMS-02 experiment operating on the International Space Station is expected to greatly improve on the current sensitivity to antideuteron fluxes~\cite{Choutko:2007}.\footnote{Note that the sensitivity in~\cite{Choutko:2007} was estimated based on 5 years of data taking in the superconducting magnet set-up of the apparatus. It is not clear whether the actual permanent magnet set-up can reach this sensitivity. To date, however, there is no updated study on the AMS-02 antideuteron sensitivity~\cite{Aramaki:2015}.} Moreover, the General AntiParticle Spectrometer (GAPS) is expected to perform several balloon flights, starting with a first Antarctic campaign in the austral summer 2019/2020~\cite{Aramaki:2015}. A prototype flight of the GAPS experiment was successfully carried out in June 2012~\cite{vonDoetinchem:2013oxa}. 

Several studies on antideuteron fluxes from DM annihilations or decays can be found in the literature~\cite{Donato:1999gy,Baer:2005tw,Donato:2008yx,Brauninger:2009pe,Ibarra:2009tn}. These early studies, however, employ the factorised coalescence approximation for antideuteron formation, which, as discussed in section 2, is in general insufficient to describe the actual production rate. Only more recent studies employ the Monte Carlo approach~\cite{Kadastik:2009ts,Cui:2010ud,Grefe:2011kh,Dal:2012my,Ibarra:2012cc,Fornengo:2013osa,Dal:2014nda,Monteux:2014tia}. As discussed in section~2, in this work we employ decay spectra obtained by this latter method.

A relativistic antideuteron, formed by the coalescence of an antineutron and an antiproton within the hadronic shower of a gravitino DM decay in the Milky Way halo, will then propagate through the ISM and might eventually arrive at a detector at Earth. As for all cosmic rays, the propagation is described by a diffusion equation of the cosmic-ray phase-space density $\psi$: 
\begin{equation}
 \begin{split}
\vec{\nabla} \cdot(\vec{V}_c\, \psi - K_0\, \beta\, \mathcal{R}^\delta\, \vec{\nabla} \psi ) +  \partial_E\left(b_{\text{loss}}\,\psi -D_{EE}\, \partial_E\psi\right) \qquad\qquad\qquad\quad \\
\qquad\qquad\quad = Q^{\text{prim}} + 2\,h\,\delta(z) \left(Q^{\text{sec}} + Q^{\text{ter}}\right) - 2\,h\,\delta(z)\,\Gamma^{\text{spal}}\,\psi.
 \end{split}
\end{equation}
For a description of the individual terms we refer to the appendix of~\cite{Delahaye:2013yqa} and references therein.

\subsection{Cross Sections}
The spallation term $\Gamma^{\text{spal}} = v_{\bar{d}}\,(n_\text{H}\sigma_{\bar{d}p}^{\text{ann}} + n_\text{He}\sigma_{\bar{d}\text{He}}^{\text{ann}})$ deserves a separate discussion. This term corresponds to the interaction of cosmic-ray antideuterons of velocity $ v_{\bar{d}}$ with the ISM, which is mainly composed of hydrogen and helium. We have considered $n_\text{H}=0.9$\,cm$^{-3}$ and $n_\text{He}=0.1$\,cm$^{-3}$~\cite{Ferriere:2007yq}. Contributions of heavier elements have been neglected. The annihilating inelastic cross section $\sigma_{\bar{d}p}^{\text{ann}} = \sigma_{\bar{d}p}^{\text{inel}} - \sigma_{\bar{d}p}^{\text{non-ann}}$ is practically given by the inelastic cross section since the non-annihilating inelastic cross section is very small due to the small binding energy of antideuterons~\cite{Duperray:2005si,Donato:2008yx,Brauninger:2009pe,Dal:2012my}. The inelastic cross section is the difference of the total and the elastic cross sections, $\sigma_{\bar{d}p}^{\text{inel}} = \sigma_{\bar{d}p}^{\text{tot}} - \sigma_{\bar{d}p}^{\text{el}}$.

A common issue in cosmic-ray antideuteron analyses is that there are no experimental data on the cross section of inelastic antideuteron--proton scattering. A typical assumption is hence to rely on the hypothesis that $\sigma^{\text{inel}}_{\bar{d}p}(T_{\bar{d}}/n) = 2\times\sigma^{\text{inel}}_{\bar{p}p}(T_{\bar{p}}=T_{\bar{d}}/n)$, where $T_{\bar{d}}/n$ is the antideuteron kinetic energy per nucleon, see~\cite{Ibarra:2012cc} using a direct parametrisation of $\sigma^{\text{inel}}_{\bar{p}p}$ data by Tan and Ng~\cite{Tan:1983de}.\footnote{Some earlier studies used the assumption $\sigma_{\bar{d}p}^{\text{ann}}(T_{\bar{d}}/n) = 2\times\sigma_{\bar{p}p}^{\text{ann}}(T_{\bar{p}})$~\cite{Ibarra:2009tn,Cui:2010ud}. This relation, however, underestimates $\sigma_{\bar{d}p}^{\text{ann}}$ since the annihilating antiproton--proton cross section falls quickly with rising energy, in contrast to the antideuteron--proton process, see appendix~A.} A similar path was followed by~\cite{Donato:2008yx}, calculating the cross section as $\sigma^{\text{inel}}_{\bar{p}d}(p_{\bar{p}}) = 2\times(\sigma^{\text{tot}}_{\bar{p}p}(p_{\bar{p}})-\sigma^{\text{el}}_{\bar{p}p}(p_{\bar{p}}))$. Making use of the reasonable assumption of symmetry under charge conjugation, one can also use the total antiproton--deuteron cross section $\sigma^{\text{tot}}_{\bar{p}d}(p_{\bar{p}})$, for which data exists~\cite{Agashe:2014kda}. In this case, the inelastic antiproton--deuteron cross section was calculated as $\sigma^{\text{inel}}_{\bar{p}d}(p_{\bar{p}}) = \sigma^{\text{tot}}_{\bar{p}d}(p_{\bar{p}})-2\times\sigma^{\text{el}}_{\bar{p}p}(p_{\bar{p}})$~\cite{Brauninger:2009pe,Dal:2012my,Fornengo:2013osa}.

However, although the assumptions presented above give the correct order of magnitude of $\sigma^{\text{inel}}_{\bar{d}p}$, they are not entirely supported by experimental evidence. In fact, $2\times\sigma^{\text{tot}}_{\bar{p}p}$ is larger than $\sigma^{\text{tot}}_{\bar{p}d}$ by roughly 10\%. This is theoretically expected due to Glauber screening~\cite{Glauber:1955qq} and in nucleon--nucleus collisions one rather expects a geometric scaling $\sigma_{pA}\simeq A^{2/3}\times\sigma_{pp}$ than a linear scaling with nucleon number $A$~\cite{Gribov:1968jf,Binon:1970yu,Mori:2009te}. 

\begin{figure}[t]
 \centering
 \includegraphics[width=0.6\linewidth]{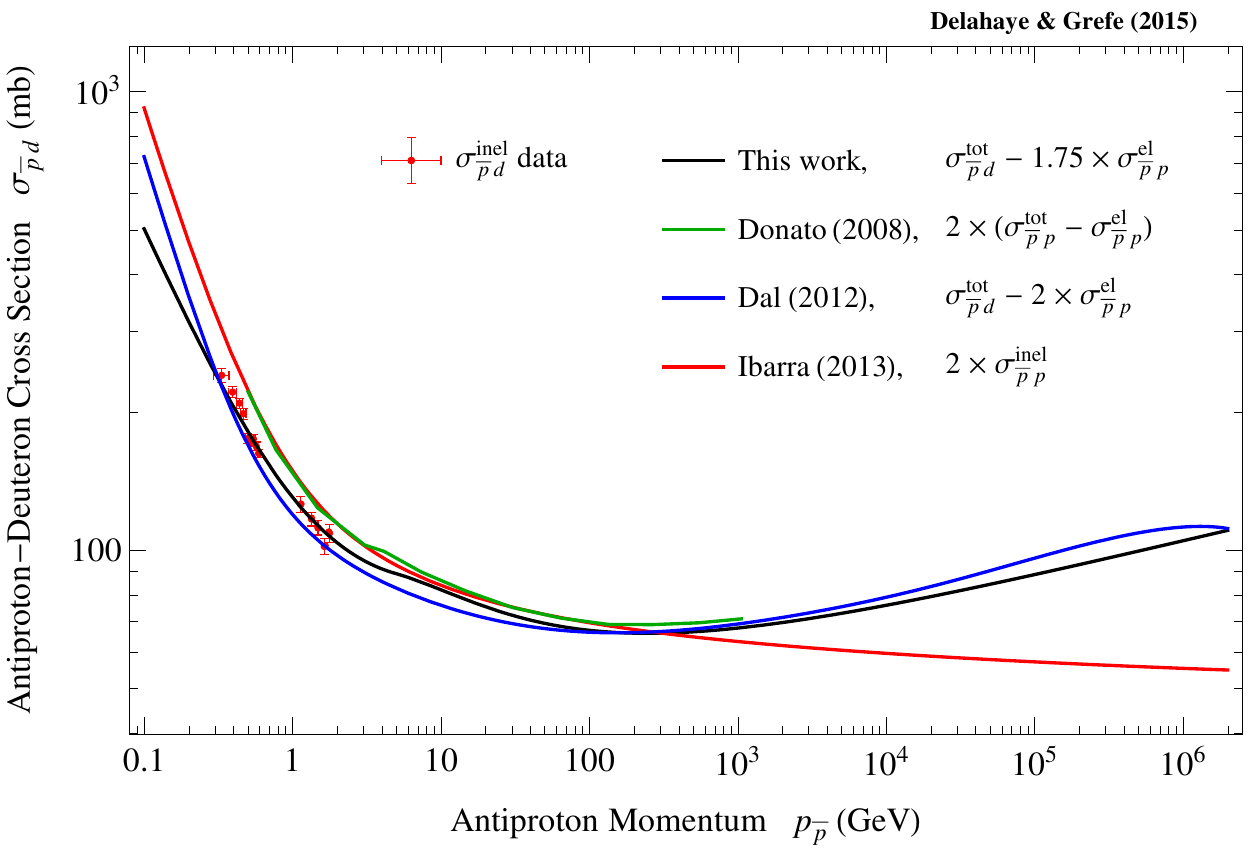} 
 \caption{Comparison of our parametrisation of the inelastic antideuteron-proton cross section to other parametrisations used throughout the literature. The data points are taken from the Landolt-B\"ornstein compilation~\cite{Baldini:1988ti}.}
 \label{CSinelastic}
\end{figure}

In this work, we will thus use an inelastic antiproton--deuteron cross section based on parametrisations of available data. Indeed, in many previous works it was incorrectly stated that there were no data on the inelastic and elastic antiproton--deuteron processes. In figure~\ref{CSinelastic}, we compare our parametrisation of $\sigma^{\text{inel}}_{\bar{p}d}$ with other parametrisations used in the literature. We observe that the approach of rescaling $\sigma^{\text{inel}}_{\bar{p}p}$ by a factor of two overshoots the available low-energy data. The parametrisation by Tan and Ng~\cite{Tan:1983de} is only based on $\sigma^{\text{inel}}_{\bar{p}p}$ data from 50\,MeV to 100\,GeV in antiproton kinetic energy and clearly leads to an incorrect high-energy behaviour. The approach of Donato \textit{et al.}~\cite{Donato:2008yx}, based on the difference of $\sigma^{\text{tot}}_{\bar{p}p}$ and $\sigma^{\text{el}}_{\bar{p}p}$ data that extend to much higher energies, leads to a more reasonable high-energy behaviour. Dal \textit{et al.}~\cite{Dal:2012my},\footnote{The used parametrisations are not given in the paper; L.~Dal, private communication (2014).} using parametrisations of $\sigma^{\text{tot}}_{\bar{p}d}$ and $\sigma^{\text{el}}_{\bar{p}p}$, find a better agreement with low-energy data. Our parametrisation gives a comparable result, but makes use of better motivated functional forms for the parametrisations and explicitly takes into account available low-energy data for the antiproton--deuteron process. See appendix~A for a detailed derivation of the parametrisation used in this work. For $\sigma_{\bar{p}d}^{\text{non-ann}}$ we use the parametrisation of Dal \textit{et al.}~\cite{Dal:2012my}. Our result is certainly not yet satisfactory, but we think that it is definitely an improvement compared to the treatment of cross sections in earlier antideuteron studies. We hope that our work serves to stimulate further discussion in this area. 

In addition to the antideuteron--proton cross section, we need the cross section for annihilating antideuteron--Helium scattering. Since no experimental data are available for this process, we have rescaled the antideuteron--proton cross sections by the geometric factor $4^{2/3}$, hoping that it suffices to give a reasonable estimate of the true cross section. 

Concerning the production of tertiaries, for which the differential cross section of the non-annihilating inelastic antideuteron--proton process is required, unfortunately no data are available. The usual method is thus to take the integrated cross section of the charge-conjugate process, $\sigma_{\bar{p}d}^{\text{non-ann}}$, and to multiply it by an appropriate energy distribution of the outgoing deuteron. In the limiting fragmentation hypothesis, this is simply $1/T'$, where $T'$ stands for the kinetic energy of the incoming deuteron (see for instance~\cite{Duperray:2005si,Tan:1983de}). However, one can also follow the authors of~\cite{Duperray:2005si,Donato:2008yx} and -- inspired by the $pp$ process -- use the functional form suggested by Anderson \textit{et al.}~\cite{Anderson:1967zzc}:
\begin{align}
\frac{d\sigma_{\bar{d}p\,\rightarrow\,\bar{d}X}}{dp}(p'\rightarrow p) &= \sigma_{\bar{p}d}^{\text{non-ann}}(p')\times\frac{d\sigma^{\text{Anderson}}}{dp} \times \left[ \int_0^{p'} \frac{d\sigma^{\text{Anderson}}}{dp''}\,dp''\right]^{-1},
\intertext{where}
\frac{d\sigma^{\text{Anderson}}}{dp} &\equiv 2\pi\int_0^{\pi}\frac{d^2\sigma(pp\rightarrow pX)}{dp\,d\Omega} \sin\theta\, d\theta\nonumber
\intertext{and}
\frac{d^2\sigma(pp\rightarrow pX)}{dp\,d\Omega} &= \frac{p^2}{2\pi p_t}\frac{\gamma(E-\beta p\cos \theta)}{E}\,610\,p_t^2 \exp\left(-\frac{p_t}{0.166}\right)\frac{\text{mb}}{\text{GeV}\,\text{sr}}.\nonumber
\end{align}
In the latter expression, the energy and momentum of the incoming proton are labelled $E$ and $p$, respectively; $p_t$ stands for the transverse momentum of the outgoing proton in units of GeV. The boost from the laboratory frame to the centre-of-mass frame is ruled by the Lorentz coefficients $\beta$ and $\gamma$. We follow this latter method in this work.

\subsection{Propagated fluxes}

\begin{figure}[t]
\centering
\includegraphics[width=0.6\linewidth]{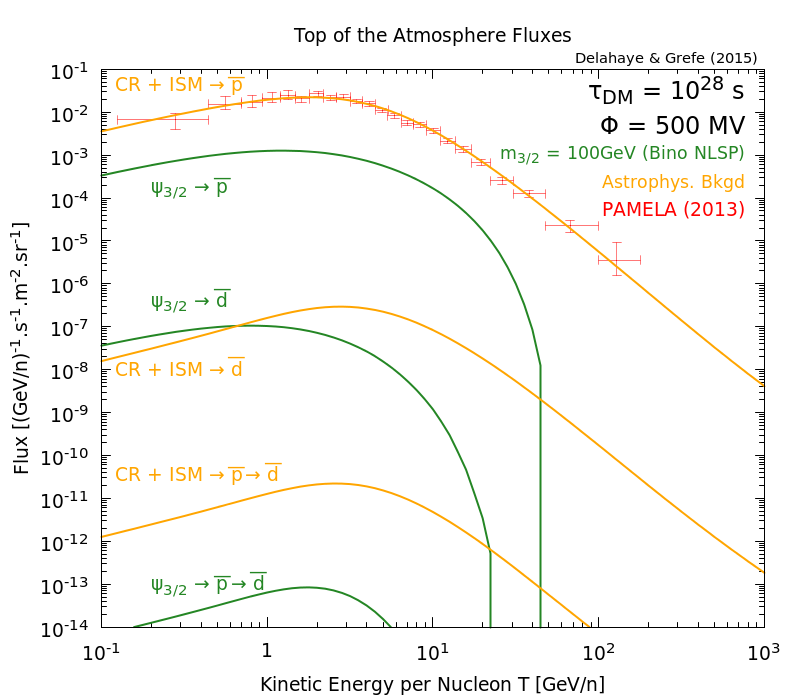} 
 \caption{Various components of the antiproton and antideuteron fluxes at Earth. The astrophysical background (secondaries) comes from the interaction of cosmic-ray protons and $\alpha$-particles with the ISM. For the case of antideuterons also interactions of secondary antiprotons with the ISM contribute. The primary antiproton and antideuteron components come from gravitino decay; for the case of antideuterons also from the interaction of primary antiprotons with the ISM. The tertiary component corresponds to a redistribution of high-energy cosmic rays to lower energies due to non-annihilating inelastic scattering off the ISM. Since every component creates tertiaries, they have been incorporated directly and are not displayed as separate components.}
\label{subfluxes}
\end{figure}

As for the case of antiprotons, no primary antideuterons are expected from astrophysical objects and the dominant background for DM searches are secondary antideuterons created in spallation processes of cosmic-ray protons and helium nuclei impinging on the ISM, i.e.\ hydrogen and helium gas. In order to estimate the background for the signal from gravitino decay, we employ here the same calculation of the astrophysical secondary antideuteron flux as used in~\cite{Donato:2008yx}. As one can see in figure~\ref{subfluxes}, various processes have to be taken into account for the calculation of the antiproton and antideuteron fluxes. The relevant processes for antiprotons are presented in table~\ref{pbarprocesses} and those for antideuterons are presented in table~\ref{dbarprocesses}. For the antiproton processes, we have used the cross section parametrisations given in~\cite{Tan:1984ha}. For the calculation of secondary antideuterons, we have used the antiproton cross sections along with the factorised coalescence prescription for antideuteron formation in the final state, see section~2.\footnote{Alternatively, one could also use the Monte Carlo approach for calculating secondaries~\cite{Ibarra:2013qt}.}
\begin{table}
  \centering
  \begin{tabular}{|lrll|}
    \hline
    Primaries & $\psi_{3/2} $& $\xrightarrow{\text{\tiny Decay}}$ & $\bar{p} + X$\\
    \hline
    \multirow{2}{*}{Secondaries}&   $p$&$ {\xrightarrow{+ \text{{\tiny ISM}}}}$&$ \bar{p} + X$ \\
    &  $\alpha$&$ {\xrightarrow{+ \text{{\tiny ISM}}}}$&$ \bar{p} + X$ \\
    \hline
    Tertiaries &  $\bar{p}$&$ {\xrightarrow{+ \text{{\tiny ISM}}}}$&$ \bar{p} + X$ \\
    \hline
\end{tabular}
\caption{Relevant processes for calculating the flux of antiprotons.\label{pbarprocesses}}
\end{table}

\begin{table}
  \centering
  \begin{tabular}{|lrll|}
    \hline
    Primaries & $\psi_{3/2} $& $\xrightarrow{\text{\tiny Decay}}$ & $\bar{d} + X$\\
    \hline
    \multirow{4}{*}{Secondaries}&   $p$&$ {\xrightarrow{+ \text{{\tiny ISM}}}}$&$ \bar{d} + X$ \\
    &  $\alpha$&$ {\xrightarrow{+ \text{{\tiny ISM}}}}$&$ \bar{d} + X$ \\
    & $\bar{p}_{\text{ Primaries}}$&$ {\xrightarrow{+ \text{{\tiny ISM}}}}$&$ \bar{d} + X$ \\
    & $\bar{p}_{\text{ Secondaries}}$&$ {\xrightarrow{+ \text{{\tiny ISM}}}}$&$ \bar{d} + X$ \\
    \hline
    Tertiaries &  $\bar{d}$&$ {\xrightarrow{+ \text{{\tiny ISM}}}}$&$ \bar{d} + X$ \\
    \hline
\end{tabular}
\caption{Relevant processes for calculating the flux of antideuterons.\label{dbarprocesses} }
\end{table}

Since both primaries and secondaries produce tertiaries, i.e.\ cosmic rays produced by the non-annihilating inelastic scattering of high-energy antideuterons on the ISM, we have not shown this component separately but rather added it directly to our estimates of the primaries and secondaries, respectively. However, we stress that this process, as well as convection and diffusive reacceleration, have important impact on the low-energy part of the computed fluxes and should not be neglected.
Note also that in the case of antideuterons the secondaries coming from the spallation of primary antiprotons, i.e.\ antiprotons of DM origin, on the ISM are to be considered as signal and not background. This component is much lower than the primary component and the background when the gravitino mass is low, but for masses higher than 10\,TeV, it is of the same order of magnitude as the primary component and hence should not be neglected. Figure~\ref{subfluxes} displays these different components for a gravitino mass of 100\,GeV. 

In order to be able to compare the expected antideuteron fluxes to experimental results, one has to take into account the effect of solar modulation~\cite{Gleeson:1968zza}. Ideally, one should try to fully model this effect, for instance with a simulation like {\sc HelioProp}~\cite{Maccione:2012cu} as done in~\cite{Fornengo:2013osa}. However, this requires knowing precisely the status of the solar environment at the time of data taking. In the absence of data and because we do not know yet the time of the data taking, we think it is untimely to go through such a precise modelling. For Figure~\ref{subfluxes}, we hence satisfy ourselves with the so-called Fisk approximation~\cite{Perko:1987}: 
\begin{equation}\label{solarmod}
\Phi^{\text{TOA}}_{\bar{d}} (E^{\text{TOA}}) = \Phi^{\text{IS}}_{\bar{d}} (E^{\text{IS}} \equiv E^{\text{TOA}}+\phi_F) \left(\frac{E^{\text{TOA}}}{E^{\text{IS}}} \right)^2\,,
\end{equation}
assuming a Fisk potential of $\phi_F=500\,$MV as an example since this values allows to have a good agreement with the most recent PAMELA antiproton data~\cite{Adriani:2012paa}. In the subsequent figures, where we do not show antiproton data, we will only consider interstellar fluxes since we do not know what the Fisk potential will be when data will finally be taken. 

\begin{figure}[t]
 \includegraphics[width=0.49\linewidth]{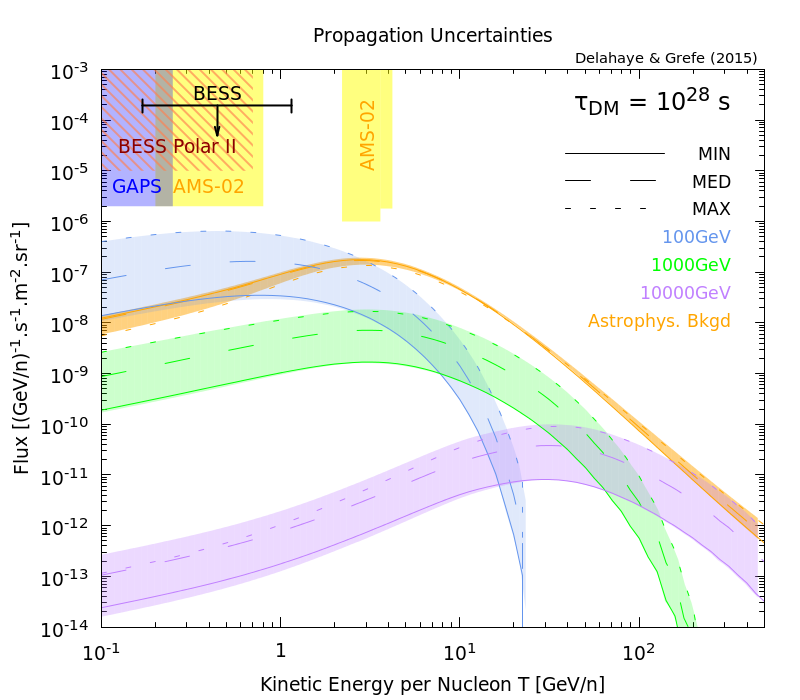} 
 \hfill
 \includegraphics[width=0.49\linewidth]{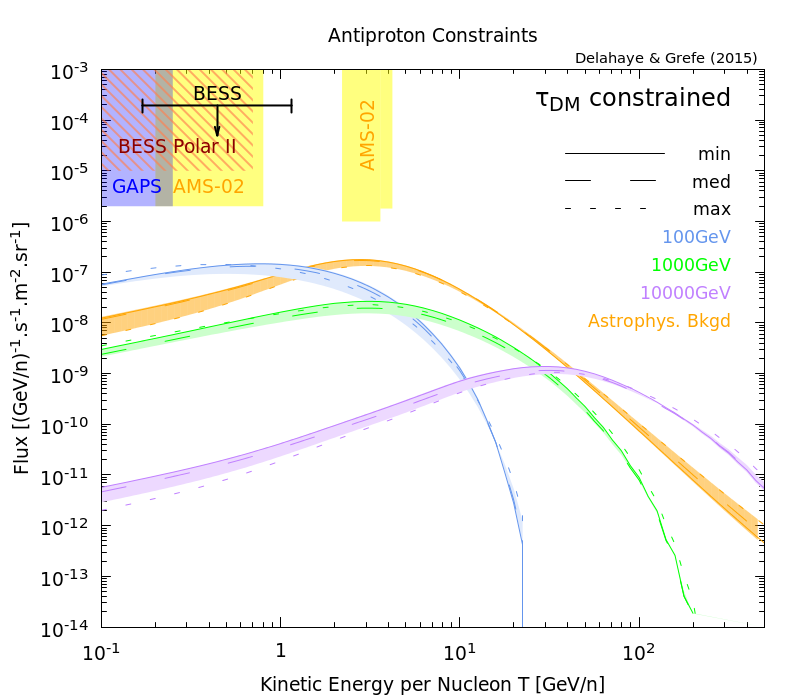} 
 \caption[Cosmic-ray antideuteron flux from gravitino decays compared to the expectation from astrophysical secondary production and the sensitivities of forthcoming experiments.]{\textit{Left:} Cosmic-ray antideuteron flux expected from the decay of gravitino DM compared to the expectation from astrophysical secondary production and the sensitivities of forthcoming experiments. The flux from gravitino decay is shown for a lifetime of $10^{28}\,$s and masses of 100\,GeV, 1\,TeV and 10\,TeV. The coloured bands correspond to propagation uncertainties within constraints from boron-to-carbon ratio measurements.
\textit{Right:} Same as left panel but fixing the gravitino lifetime to the lowest value allowed by antiproton constraints.
No solar modulation has been implemented here since the time of data taking is unknown.}
 \label{antidflux}
\end{figure}
In figure~\ref{antidflux}, we present the interstellar antideuteron spectrum from gravitino DM decays and compare it to the expected astrophysical background and the flux limit obtained by the BESS experiment~\cite{Fuke:2005it}. In addition, we present the projected sensitivity regions of the BESS-Polar II~\cite{Sasaki:2014}, AMS-02~\cite{Choutko:2007} and GAPS~\cite{vonDoetinchem:2013oxa}\footnote{The GAPS sensitivity assumed in this work corresponds to three Antarctic long duration balloon flights with a total duration of 105 days~\cite{Aramaki:2015}.} experiments. In the left panel, we show the uncertainty band due to the lack of precise knowledge of the propagation parameters as constrained by measurements of the boron-to-carbon ratio in cosmic rays~\cite{Maurin:2001sj}. The coloured bands correspond to a full scan over the allowed parameter space. As an illustration we also display the fluxes obtained with three benchmark models often used in the literature (see table~\ref{antideuteronparameters}). We use a common decay lifetime of $\tau_{3/2} = 10^{28}$\,s for illustration. Note that, as for the case of antiprotons, the MIN/MED/MAX benchmark models do not size the full extent of the uncertainty band. This shows again the importance of performing scans over the full propagation parameter space allowed by other data rather than checking only a few cases.
\begin{table}[t]
 \centering
 \begin{tabular}{ccccc}
  \toprule
  Model & $\delta$ & $K_0\,(\text{kpc}^2\!/\text{Myr})$ & $L\,(\text{kpc})$ & $V_C\,(\text{km/s})$ \\
  \midrule
  MIN & 0.85 & 0.0016 & 1 & 13.5 \\
  MED & 0.70 & 0.0112 & 4 & 12 \\
  MAX & 0.46 & 0.0765 & 15 & 5 \\
  \bottomrule
 \end{tabular}
 \caption{Parameters of cosmic-ray propagation models that correspond, respectively, to the best fit of cosmic-ray boron-to-carbon data (MED) as well as the minimal (MIN) or maximal (MAX) antiproton flux compatible with cosmic-ray boron-to-carbon data. Figures taken from~\cite{Donato:2003xg}.}
 \label{antideuteronparameters}
\end{table}

In the right panel of figure~\ref{antidflux}, we display the same fluxes but setting the decay lifetime to the minimum values allowed by the constraints obtained using the antiproton measurements (see~\cite{Delahaye:2013yqa}). Note that the coloured bands correspond to the extreme cases obtained for antiprotons and are not the full uncertainty band. Indeed, the limiting cases correspond to situations where the antiproton flux is maximised at a given energy bin. This does not mean that the antideuteron flux is maximal over the whole energy range.

\begin{figure}[t]
\centering
 \includegraphics[width=0.6\linewidth]{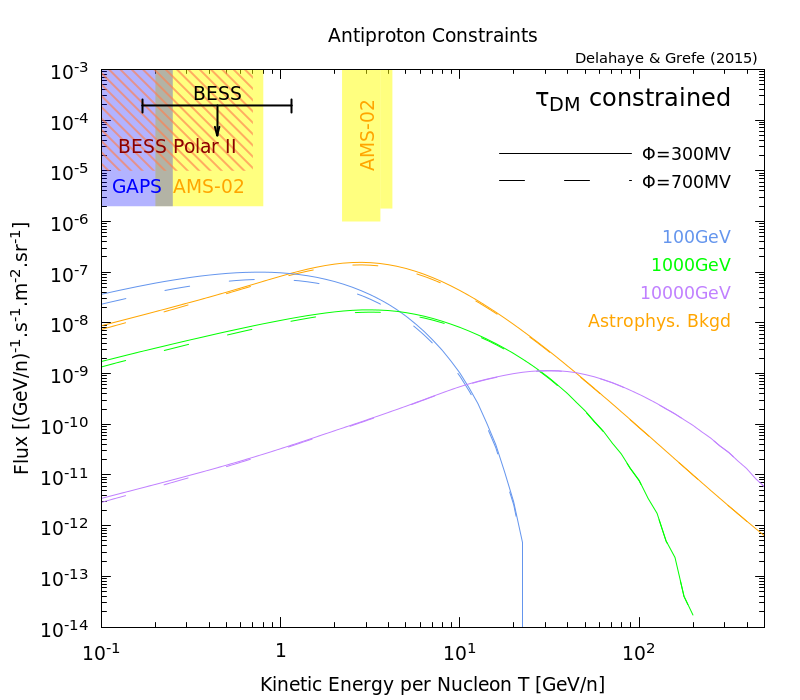} 
 \caption[Effect of Solar modulation.]{Same as the right panel of Fig.~\ref{antidflux} for the MED propagation parameters. The only difference is that the fluxes displayed are now corrected for solar modulation ($\phi_F=300\,$MV and $\phi_F=700\,$\,MV). As one can see, because the fluxes are relatively flat below 10\,GeV, solar modulation has little impact on the expected fluxes.}
 \label{solmod}
\end{figure}
As one can see from Fig.~\ref{solmod}, if the Fisk approximation describes correctly the influence of solar modulation on antideuterons, then this should not affect our conclusions dramatically. Indeed, unlike many other cosmic-ray species, antideuteron fluxes are relatively flat below 10\,GeV and a shift of the antideuteron energy does not affect the flux very strongly, at least within a reasonable range of the Fisk potential.

\section{Discussion of the Detection Prospects}

A striking feature of figure~\ref{antidflux} is that for masses as low as 100~GeV the gravitino decay signal can be of the same order as the astrophysical background below a few GeV, even for lifetimes as large as $10^{28}\,$s, a value not yet excluded by gamma-ray and antiproton observations (see for instance~\cite{Grefe:2014bta}).\footnote{See also~\cite{Carquin:2015uma,Ando:2015qda} for recent works on gamma-ray constraints on the gravitino lifetime, taking into account the latest Fermi LAT data~\cite{Ackermann:2014usa}.} In this respect, it would also be interesting to see what antideuteron flux could be expected for even lower gravitino masses. It could thus be worthwhile to study this region in a future work using the spectra obtained from gravitino three-body decays~\cite{Grefe:2011dp}.

But also for larger gravitino masses the antideuteron signal could be at the same order as the background. This was not observed in earlier studies as the signal for large DM masses is artificially suppressed in the factorised coalescence prescription. Therefore, for decaying DM candidates there is in principle also the possibility of observing an exotic component in the higher-energetic part of the spectrum, where currently no experiments are planned. Note, however, that both background and signal are extremely low and quite challenging for experimentalists as this would mean improving sensitivity by at least four orders of magnitude in flux, but also to reach much higher energies. 

When taking into account the constraints derived from antiproton observations~\cite{Delahaye:2013yqa}, we find that the remaining parameter space for having a gravitino decay signal significantly higher than the astrophysical background becomes quite small but does not vanish completely. Since the coalescence process still suffers large theoretical uncertainties, one cannot exclude that all the fluxes are in fact larger (or smaller) than what we assume here. The 1-$\sigma$ uncertainty in $p_0$ could lead to an increase of roughly 30\% in the antideuteron flux from gravitino decay, irrespective of the antiproton constraints. In addition, making use of Monte Carlo methods to estimate the production of secondary antideuterons instead of the factorised coalescence model used here, tends to predict a slightly lower background level at low energies~\cite{Ibarra:2013qt}. This would increase the signal-to-background ratio, independently of any other constraints.

Still, it seems clear that the current generation of experiments will not be able to observe any antideuteron events. Only an improvement of the flux sensitivity by two orders of magnitude should at least allow for a detection of astrophysical antideuterons -- or even those coming from gravitino decay. The main hopes seem to reside either in the highest energy range (above $\sim50\,$GeV) or in the lowest one (below $\sim1\,$GeV). Note, however, that the latter is affected by solar modulation, a phenomenon that to date is not fully under modelling control.

This clearly challenges antideuterons as the golden channel it has long thought to be. Indeed the antiproton channel has become extremely constraining thanks to the PAMELA data~\cite{Adriani:2012paa}. A forthcoming release of AMS-02 antiproton data could make these constraints a bit more stringent, especially at higher energies.

\section{Conclusion}

In this work, we have studied the potential of detecting cosmic-ray antideuterons produced in the decay of gravitino dark matter within a framework of bilinear $R$-parity violation. This work was a natural sequel of a related work concerning cosmic-ray antiprotons. After discussing the deuteron formation in hadronic showers and calculating antideuteron spectra from gravitino decay, we have determined the gravitino decay signal at Earth. We have also assessed the uncertainties affecting the expectations for the astrophysical background and the signal. We have shown that there is some room left for a discovery of gravitino decays through antideuterons, however not within the sensitivity of the current and planned generation of experiments.

We have also shown that -- once the antiproton constraints are taken into account -- the remaining parameter space for a detection of gravitino dark matter with bilinear $R$-parity violation via antideuterons is quite small. On the other hand, since not much progress is expected in the background-limited antiproton channel in the coming years, the antideuteron channel could still serve to put stronger constraints on the strength of the $R$-parity violation in the future. If the detection technology were to improve considerably, also the high-energy regime (above $\sim50\,$GeV) would become interesting for the search of gravitino dark matter.

\section*{Acknowledgements}
MG is grateful to Philip von Doetinchem and Rene Ong for the organisation of the Antideuteron 2014 workshop and thus providing a stimulating environment for discussions on antideuterons among experts in the field. MG would also like to thank Nicolao Fornengo, Sebastian Wild and Andrea Vittino for discussions on the coalescence prescription, and Lars A.\ Dal for discussions on the cross sections relevant for antideuteron propagation.

The work of TD was supported by the Swedish Research Council under contract number 349-2007-8709. TD is also grateful for support by the Forschungs- und Wissenschaftsstiftung Hamburg through the programme ``Astroparticle Physics with Multiple Messengers''.
The work of MG was supported by the Forschungs- und Wissenschaftsstiftung Hamburg through the programme ``Astroparticle Physics with Multiple Messengers'', by the Marie Curie ITN ``UNILHC'' under grant number PITN-GA-2009-237920 and by the Spanish MINECO's ``Centro de Excelencia Severo Ochoa'' programme under grant SEV-2012-0249. MG is also grateful for support by the Marie Curie ITN ``INVISIBLES'' under grant number PITN-GA-2011-289442.

\begin{appendix}

\section{Cross Sections}
In this appendix, we briefly present the parametrisations for the antiproton--proton and antiproton--deuteron cross sections we used to estimate the inelastic scattering cross sections relevant for the spallation term of antideuteron propagation in the Milky Way.

\paragraph{Antiproton--Proton Cross Sections} 
For the total cross section of antiproton--proton scattering, a large number of data points ranging from roughly 200\,MeV to 2\,PeV in antiproton momentum (in the rest frame of the proton target) exist in the literature~\cite{Agashe:2014kda}. A useful parametrisation is given by~\cite{Arkhipov:1999yw,Arkhipov:1999sy}:
\begin{align}
\sigma_{\bar{p}p}^\text{tot}(s) &= \sigma_\text{asmpt}^\text{tot}(s)\left[1+\frac{c}{\sqrt{s-4\,m_p^2}R_0^3(s)}\left(1+\frac{d_1}{\sqrt{s}}+\frac{d_2}{s}+\frac{d_3}{s^{3/2}}\right)\right],
\label{pbarpArkhipov}
\end{align}
where $s = 2\,m_p^2 + 2\,m_p\sqrt{m_p^2 + p_{\bar{p}}^2}$ is the centre-of-mass energy of the $\bar{p}p$ system and
\begin{align*}
\sigma_\text{asmpt}^\text{tot}(s) &= a_0+a_2\ln^2(\sqrt{s}/\sqrt{s_0})\,, & 
\sqrt{s_0} &= (20.74\pm1.21)\,\text{GeV}, \\
R_0^2(s) &= 2.568\,\text{GeV}^{-2}\,\text{mb}^{-1}\,\frac{\sigma_\text{asmpt}^\text{tot}(s)}{2\pi} - B(s)\,, &
B(s) &= b_0+b_2\ln^2(\sqrt{s}/\sqrt{s_0}).
\end{align*}
In the definition of $R_0$, the factor $1\,\text{mb} = 2.568\,\text{GeV}^{-2}$ simply accounts for unit conversion. The remaining parameters in these expressions were determined from a fit to experimental data in~\cite{Arkhipov:1999yw}:
\begin{align*}
a_0 &= (42.05\pm0.11)\,\text{mb}, & a_2 &= (1.755\pm0.083)\,\text{mb}, \\ 
b_0 &= (11.92\pm0.15)\,\text{GeV}^{-2}, & b_2 &= (0.304\pm0.019)\,\text{GeV}^{-2}, & c &= (6.7\pm1.8)\,\text{GeV}^{-2}, \\ 
d_1 &= (-12.1\pm1.0)\,\text{GeV}, & d_2 &= (90\pm16)\,\text{GeV}^2, & d_3 &= (-111\pm22)\,\text{GeV}^3.
\end{align*}
With these parameters, eq.~(\ref{pbarpArkhipov}) gives a relatively good fit to the available data. Including also high-energy proton--proton data between 10\,TeV and 2\,EeV in proton momentum -- where the antiproton--proton and proton--proton cross sections should be equivalent -- we get a goodness of fit of $\chi^2/\text{dof} = 2821/460$.\footnote{We have also checked that the total cross section data for the antiproton--neutron~\cite{Agashe:2014kda} and antineutron--proton~\cite{Baldini:1988ti} scattering processes are compatible with the antiproton--proton result.} Given that many early experimental works, especially from the 1960s and 1970s, only quote statistical errors without any assessment of systematic uncertainties, the fit is quite acceptable (see also the remark on errors in the introduction of~\cite{Baldini:1988ti}).

For the elastic cross section of antiproton--proton scattering, there is also a lot of data ranging from roughly 200\,MeV to 2\,PeV in antiproton momentum~\cite{Agashe:2014kda}. Using typical basis functions for cross sections (see~\cite{Baldini:1988ti}), we fit the data using the piecewise ansatz:\footnote{The authors of~\cite{Uzhinsky:2011zz} present a parametrisation of the elastic antiproton--proton cross section based on the methods of~\cite{Arkhipov:1999yw,Arkhipov:1999sy}. However, it appears that their parameters are faulty since we were not able to reproduce their result.}
\begin{equation}
\sigma_{\bar{p}p}^\text{el}(p_{\bar{p}}) = \begin{cases} a_0 + a_1 (p_{\bar{p}}/\text{GeV})^n &\text{if } p_{\bar{p}} \leq 5\,\text{GeV} \\
b_0 + b_1 (p_{\bar{p}}/\text{GeV})^m + b_2 \ln (p_{\bar{p}}/\text{GeV}) + b_3 \ln^2 (p_{\bar{p}}/\text{GeV})  &\text{if } p_{\bar{p}} > 5\,\text{GeV} \end{cases}
\label{pbarpelastic}
\end{equation}
and requiring differentiability at $p_{\bar{p}} = 5\,$GeV. We find the parameters 
\begin{align*}
a_0 &= -35.71\,\text{mb}, & a_1 &= 81.03\,\text{mb}, & n &= -0.2567, & b_0 &= 10.49\,\text{mb}, \\
b_1 &= 86.04\,\text{mb}, & m &= -1.385, & b_2 &= -1.360\,\text{mb}, & b_3 &= 0.1312\,\text{mb},
\end{align*}
giving a goodness of fit of $\chi^2/\text{dof} = 364.6/167$.\footnote{We have also checked that the cross section data for the elastic antiproton--neutron~\cite{Agashe:2014kda} and antineutron--proton~\cite{Baldini:1988ti} scattering processes are compatible with the antiproton--proton result.} Besides data on elastic antiproton--proton scattering we included data on elastic proton--proton scattering with incoming momentum between 100\,GeV and 34\,PeV~\cite{Agashe:2014kda}.

For the inelastic antiproton--proton cross section there is considerably less data than in the previous cases, ranging only from 300\,MeV to 175\,GeV~\cite{Baldini:1988ti}. This cross section is, by definition, the difference between the total and the elastic cross section and we just check the consistency with available data. Including also data on inelastic proton--proton scattering between 1 and 2\,TeV, we find a goodness of fit of $\chi^2/\text{dof} = 404/43$. Given that only three of the 43 data points come with an estimate of systematic uncertainties, we think this is an acceptable result.\footnote{We have also checked that the inelastic antiproton--neutron scattering data~\cite{Baldini:1988ti} are compatible with the antiproton--proton result.}

\begin{figure}[t]
 \includegraphics[width=0.49\linewidth]{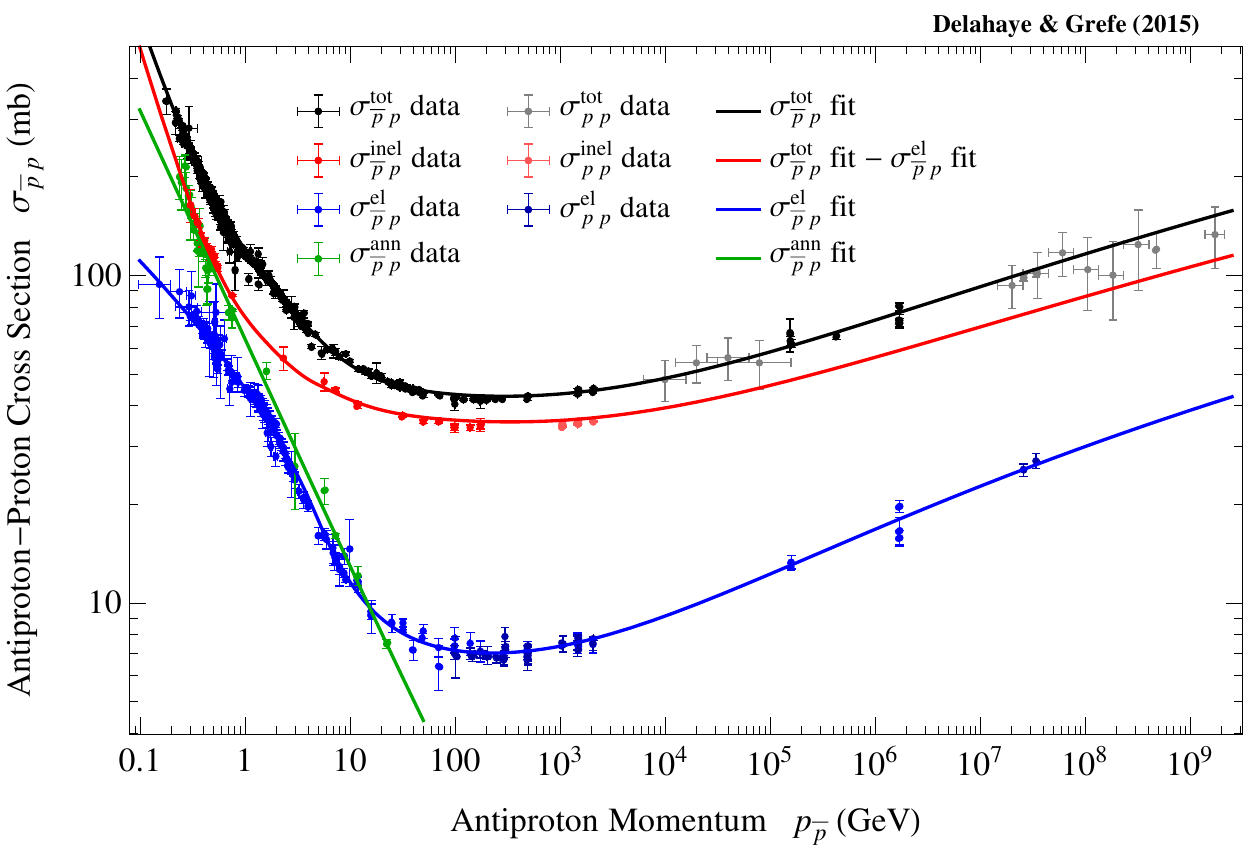} 
 \hfill
 \includegraphics[width=0.49\linewidth]{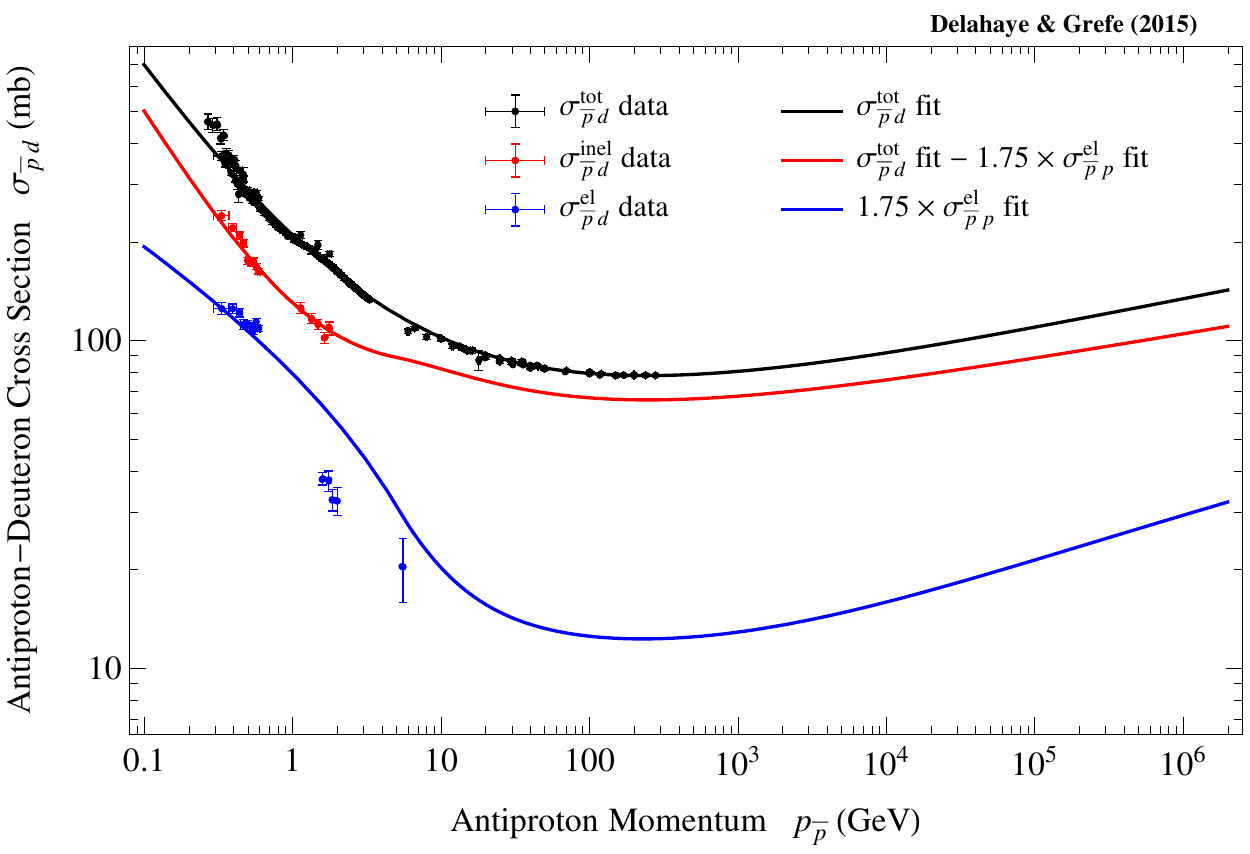} 
 \caption[Cross-sections.]{\textit{Left:} Total, elastic, inelastic and annihilating inelastic cross sections of the $\bar{p}p$ process. Data points are taken from the Particle Data Group~\cite{Agashe:2014kda} and the Landolt--B\"ornstein compilation~\cite{Baldini:1988ti}. The lines are the parametrisations discussed in the text. \textit{Right:} Same as left panel but for the cross sections of the $\bar{p}d$ process.}
 \label{cross-sections}
\end{figure}

An overview of the antiproton--proton cross section data is presented in figure~\ref{cross-sections}. In addition to the total, elastic and inelastic cross sections, we present data on annihilating inelastic antiproton--proton scattering ranging from 240\,MeV to 22\,GeV~\cite{Baldini:1988ti}. These data can be parametrised by the function
\begin{equation}
\sigma_{\bar{p}p}^\text{ann}(p_{\bar{p}}) = 0.2367\,\text{mb} + 63.86\,\text{mb}\,(p_{\bar{p}}/\text{GeV})^{-0.699},
\end{equation}
giving a goodness of fit of $\chi^2/\text{dof} = 46.5/27$.

\paragraph{Antiproton--Deuteron Cross Sections} 
The total cross section for antiproton--deuteron scattering is almost twice as large as the total antiproton--proton cross section. The exact result depends on the size of the elastic and inelastic Glauber shadow or screening corrections~\cite{Glauber:1955qq,Franco:1965wi,Arkhipov:2000hg}:
\begin{equation*}
\sigma_{\bar{p}d}^\text{tot} = 2\,\sigma_{\bar{p}p}^\text{tot} - \delta^\text{el} - \delta^\text{inel}.
\end{equation*}
If these correction terms are known, also the elastic and inelastic cross section components can be easily determined individually:
 \begin{equation*}
\sigma_{\bar{p}d}^\text{el} = 2\,\sigma_{\bar{p}p}^\text{el} - \delta^\text{el},\qquad\sigma_{\bar{p}d}^\text{inel} = 2\,\sigma_{\bar{p}p}^\text{inel} - \delta^\text{inel}.
\end{equation*}
Arkhipov describes antiproton--deuteron scattering and the corresponding correction terms in~\cite{Arkhipov:2000hg}. Unfortunately, his parametrisations do not describe very well the available data. Therefore, we will follow an alternative route.

For the total cross section of antiproton--deuteron scattering, there is only data from roughly 300\,MeV to 280\,GeV in antiproton momentum~\cite{Agashe:2014kda}. For the high-energy part we thus assume that the cross section matches the antiproton--proton cross section, rescaled with a suitable factor. We find that eq.~(\ref{pbarpArkhipov}), rescaled with a factor of 1.85, gives a reasonably good fit to the antiproton--deuteron data above 6\,GeV. Using the same basis functions as for the elastic antiproton--proton cross section, we then fit the data using the piecewise ansatz:
\begin{equation}
\sigma_{\bar{p}d}^\text{tot}(p_{\bar{p}}) = \begin{cases} a_0 + a_1 (p_{\bar{p}}/\text{GeV})^n &\text{if } p_{\bar{p}} \leq 1.35\,\text{GeV} \\
b_0 + b_1 (p_{\bar{p}}/\text{GeV})^m + b_2 \ln (p_{\bar{p}}/\text{GeV}) + b_3 \ln^2 (p_{\bar{p}}/\text{GeV})  &\text{if } p_{\bar{p}} > 1.35\,\text{GeV} \end{cases}
\end{equation}
and requiring differentiability at $p_{\bar{p}} = 1.35\,$GeV. Including rescaled antiproton--proton data above 1\,TeV and rescaled proton--proton data above 10\,TeV in the fit, we find 
\begin{align*}
a_0 &= 75.59\,\text{mb}, & a_1 &= 134.3\,\text{mb}, & n &= -0.6647, & b_0 &= 77.33\,\text{mb}, \\
b_1 &= 132.4\,\text{mb}, & m &= -0.6388\,, & b_2 &= -3.638\,\text{mb}, & b_3 &= 0.5615\,\text{mb},
\end{align*}
 giving a goodness of fit of $\chi^2/\text{dof} = 452.9/159$.

For inelastic antiproton--deuteron scattering, there are data from roughly 300\,MeV to 2\,GeV~\cite{Baldini:1988ti}. As for the case of antiproton--proton scattering, we will determine the inelastic cross section by subtracting the elastic from the total cross section. Lacking a physically motivated ansatz for the elastic antiproton--deuteron cross section, we rescale eq.~(\ref{pbarpelastic}) by a suitable factor to match the data points. We find that
\begin{equation}
\sigma_{\bar{p}d}^\text{inel}(p_{\bar{p}}) = \sigma_{\bar{p}d}^\text{tot}(p_{\bar{p}}) - 1.75 \times\sigma_{\bar{p}p}^\text{el}(p_{\bar{p}})
\end{equation}
gives a goodness of fit of $\chi^2/\text{dof} = 15.8/13$. We still have to check if our approach for the elastic cross section matches the available data ranging from roughly 300\,MeV to 5\,GeV~\cite{Baldini:1988ti}. As can be seen from the right panel of figure~\ref{cross-sections}, the shape of the rescaled elastic cross section does not entirely follow the data points. The goodness of fit thus is a very poor $\chi^2/\text{dof} = 535/9$.\footnote{Note that the agreement for the elastic cross section would be even much worse if we had determined the inelastic antiproton--deuteron cross section by rescaling the inelastic antiproton--proton cross section.} Considering only the low-energy data would result in an acceptable $\chi^2/\text{dof} = 46/9$. 

For the non-annihilating inelastic antiproton--deuteron cross section that is necessary for the calculation of tertiaries we use the parametrisation of~\cite{Dal:2012my}:\footnote{The parametrisation is not given in the paper; L.~Dal, private communication (2014 and 2015).}
\begin{equation}
\sigma_{\bar{p}d}^\text{non-ann}(p_{\bar{p}}) = 10^{-2.141 + 5.865\exp(-\log_{10}(p_{\bar{p}}/\text{GeV})) - 3.398\exp(-2\log_{10}(p_{\bar{p}}/\text{GeV}))}.
\end{equation}

\end{appendix}

\end{document}